\documentclass[12pt]{article}
\usepackage{amsmath,amssymb,amsthm,amsxtra,overpic,bbm,bm,epsfig,ulem}
\usepackage{color}
\usepackage{cite}

\usepackage{hyperref}
\usepackage{url}
\usepackage{multirow}
\usepackage{textcomp}

\def\thefootnote{\fnsymbol{footnote}}

\begin{document}

\vspace{0.2cm}

\begin{center}
{\large\bf On the two-loop radiative origin of the smallest neutrino mass \\
and the associated Majorana CP phase}
\end{center}

\vspace{0.2cm}

\begin{center}
{\bf Zhi-zhong Xing$^{1,2}$}
\footnote{E-mail: xingzz@ihep.ac.cn}
and {\bf Di Zhang$^{1}$}
\footnote{E-mail: zhangdi@ihep.ac.cn (corresponding author)}
\\
{\small $^{1}$Institute of High Energy Physics and School of Physical Sciences, \\
University of Chinese Academy of Sciences, Beijing 100049, China \\
$^{2}$Center of High Energy Physics, Peking University, Beijing 100871, China}
\end{center}

\vspace{2cm}
\begin{abstract}
Given a massless neutrino at a superhigh energy scale $\Lambda$ (e.g., in the minimal
seesaw model with only two heavy Majorana neutrinos), we calculate quantum corrections
to its initially vanishing mass $m^{}_1$ (or $m^{}_3$) and the associated Majorana CP
phase $\rho$ (or $\varrho$) at the Fermi scale $\Lambda^{}_{\rm F}$ by means of the
two-loop renormalization-group equations (RGEs) in the standard model and with the
help of the latest neutrino oscillation data. The numerical results obtained from
our analytical approximations are in good agreement with those achieved by numerically
solving the two-loop RGEs. In particular, we confirm that a nonzero value of
$m^{}_1$ (or $m^{}_3$) of ${\cal O}(10^{-13})$ eV at $\Lambda^{}_{\rm F}$
can be radiatively generated from $m^{}_1 =0$ (or $m^{}_3 =0$) at
$\Lambda \simeq 10^{14}$ GeV in the SM, and find that $\rho$ (or $\varrho$)
may accordingly acquire an appreciable physical value. As a nontrivial by-product,
the evolution of all the other (initially nonzero) flavor parameters of massive
neutrinos is studied both analytically and
numerically, by just keeping their leading (i.e., one-loop) RGE-induced effects.
\end{abstract}

\newpage

\def\thefootnote{\arabic{footnote}}
\setcounter{footnote}{0}
\setcounter{figure}{0}

\section{Introduction}

One of the most important tasks in neutrino physics and cosmology is to
determine the absolute neutrino mass scale or, equivalently, to tell how
small the smallest neutrino mass is. From a phenomenological point of view,
the lightest neutrino is allowed to be massless because this expectation
is not in conflict with current neutrino oscillation data and cosmological
observations \cite{Tanabashi:2018oca}. On the theoretical side, however,
there is no fundamental symmetry or conservation law to protect a massless
neutrino to stay massless, and hence it is most likely to become massive
after proper quantum corrections are taken into account \cite{Cai:2017jrq}.

To generate finite but tiny neutrino masses, one may extend the standard
model (SM) of electroweak interactions by adding three heavy (right-handed)
neutrino fields $N^{}_{\alpha \rm R}$ (for $\alpha = e, \mu, \tau$) and allowing
lepton number violation. In this case the charged-lepton and neutrino mass
terms that respect the $\rm SU(2)^{}_{\rm L} \times U(1)^{}_{\rm Y}$ gauge
symmetry can be written as
\begin{eqnarray}
-{\cal L}^{}_{\rm lepton} = \overline{\ell^{}_{\rm L}}
Y^{}_l H E^{}_{\rm R} + \overline{\ell^{}_{\rm L}} Y^{}_\nu
\widetilde{H} N^{}_{\rm R} + \frac{1}{2} \overline{N^{c}_{\rm R}}
M^{}_{\rm R} N^{}_{\rm R} + {\rm h.c.} \; ,
\end{eqnarray}
in which the relevant field notations are self-explanatory, and $M^{}_{\rm R}$
is a symmetric matrix whose mass scale can be far above the Fermi scale
$\Lambda^{}_{\rm F} \sim 10^2$ GeV. Integrating out the heavy degrees of freedom
in Eq.~(1) \cite{Xing:2011zza}, one is left with the unique dimension-five
Weinberg operator
\cite{Weinberg:1979sa}
\begin{eqnarray}
{\cal O}^{}_{\rm Weinberg} = \frac{\kappa^{}_{\alpha\beta}}{2} \left[
\overline{\ell^{}_{\alpha \rm L}} \widetilde{H} \widetilde{H}^T
\ell^{c}_{\beta \rm L} \right] \;
\end{eqnarray}
with the subscripts $\alpha$ and $\beta$ running over $e$, $\mu$ and $\tau$,
and the effective neutrino coupling matrix $\kappa = Y^{}_\nu M^{-1}_{\rm R} Y^T_\nu$
is suppressed by a sufficiently high cut-off scale $\Lambda$
\cite{Minkowski:1977sc,Yanagida:1979as,GellMann:1980vs,Glashow:1979nm,
Mohapatra:1979ia}. Once the electroweak gauge symmetry is spontaneously
broken at the Fermi scale $\Lambda^{}_{\rm F}$, we arrive at the effective
Majorana neutrino mass matrix for three light (left-handed) neutrinos:
\begin{eqnarray}
M^{}_\nu = - \kappa \langle H \rangle^2 = - M^{}_{\rm D} M^{-1}_{\rm R} M^T_{\rm D} \;
\end{eqnarray}
with $M^{}_{\rm D} = Y^{}_\nu \langle H\rangle$ and the charged-lepton mass
matrix $M^{}_l = Y^{}_l \langle H\rangle$, where $\langle H\rangle \simeq 174$ GeV
is the vacuum expectation value of the Higgs field. The tiny neutrino masses
$m^{}_i$ (for $i = 1, 2, 3$), which equal the singular values of $M^{}_\nu$,
are therefore ascribed to the huge mass scale of $M^{}_{\rm R}$ as compared with
the value of $\langle H\rangle$.

Eq.~(3) tells us that one of the three light neutrinos is naturally massless in
the minimal type-I seesaw scenario with only two heavy Majorana neutrinos
\cite{Kleppe:1995zz,Ma:1998zg,Frampton:2002qc}, simply because in this case the
rank of $M^{}_\nu$ is exactly equal to two (i.e., the rank of the $2\times 2$ mass
matrix $M^{}_{\rm R}$). Combining this observation with current neutrino oscillation
data \cite{Tanabashi:2018oca,Capozzi:2018ubv,Esteban:2018azc,Capozzi:2020qhw}, one
may have either $m^{}_1 =0$ (normal mass ordering) or $m^{}_3 =0$ (inverted mass
ordering). Note that the vanishing of $m^{}_1$ (or $m^{}_3$) allows one of the
Majorana CP phases in the $3\times 3$ Pontecorvo-Maki-Nakagawa-Sakata (PMNS) neutrino
mixing matrix $U$ \cite{Pontecorvo:1957cp,Maki:1962mu,Pontecorvo:1967fh}, which is used
to diagonalize $M^{}_\nu$ in the $Y^{}_l = {\rm Diag}\{y^{}_e, y^{}_\mu, y^{}_\tau\}$
basis (i.e., $U^\dagger M^{}_\nu U^* = D^{}_\nu \equiv
{\rm Diag}\{m^{}_1, m^{}_2, m^{}_3\}$ in this
basis), to automatically disappear. Such a simplified seesaw scenario is therefore
more predictive \cite{Guo:2006qa}. Of course, assuming $m^{}_1 =0$
(or $m^{}_3 =0$) and studying its phenomenological consequences are unnecessarily
subject to the minimal seesaw model, since such a conjecture empirically satisfies
the principle of Occam's razor \cite{Xing:2019vks}. Here the main concerns are
as follows: (1) whether $m^{}_1 =0$ (or $m^{}_3 =0$) can be stable against
quantum corrections between a superhigh cut-off (or seesaw) scale and the electroweak
scale; (2) whether the initially undefined Majorana CP phase $\rho$ (or $\varrho$) can
be radiatively generated together with $m^{}_1$ (or $m^{}_3$); and (3) how those
initially nonzero flavor parameters are modified by the relevant quantum effects.

The first question has essentially been answered by Davidson, Isidori and Strumia
\cite{Davidson:2006tg}. Given $m^{}_{\rm min} =0$ with $m^{}_{\rm min}$ being
either $m^{}_1$ or $m^{}_3$ at a superhigh energy scale $\Lambda \simeq 10^{14}$ GeV,
they found $m^{}_{\rm min} \sim 10^{-13}$ eV at the Fermi scale $\Lambda^{}_{\rm F}$ by
considering the two-loop renormalization-group equations (RGEs) of $M^{}_\nu$ and inputting
the preliminary neutrino oscillation data obtained in 2007. Although the Majorana
CP phase associated with $m^{}_{\rm min}$ was also mentioned in their paper, it was
not analytically formulated and numerically evaluated. On the other hand, it is
certainly enough to calculate the one-loop RGE-induced quantum corrections to those
initially nonzero flavor parameters \cite{Mei:2003gn}, but a transparent
analytical formulation of their running effects between $\Lambda^{}_{\rm F}$ and
$\Lambda \gg \Lambda^{}_{\rm F}$ has been lacking.

In this paper we are going to answer the above three questions by means of the
two-loop RGEs and with the help of the latest neutrino oscillation data in the SM
framework. Different from the previous work done by Davidson {\it et al} in
Ref.~\cite{Davidson:2006tg}, here both the smallest neutrino mass ($m^{}_1$
or $m^{}_3$) and the associated Majorana CP phase $\rho$ (or $\varrho$) at low energies
are analytically formulated by keeping the contributions of all the three neutrino
mixing angles, and their magnitudes are evaluated both based on our analytical
approximations and by numerically solving the two-loop RGEs.
The numerical results obtained in these two ways are in good agreement with
each other. In particular, we confirm that a nonzero value of
$m^{}_1$ (or $m^{}_3$) of ${\cal O}(10^{-13})$ eV at $\Lambda^{}_{\rm F}$
can be radiatively generated from $m^{}_1 =0$ (or $m^{}_3 =0$) at
$\Lambda \simeq 10^{14}$ GeV in the SM, and find that $\rho$ (or $\varrho$)
may accordingly acquire an appreciable physical value. As a nontrivial by-product,
the running behaviors of all the other (initially nonzero) flavor parameters of
massive neutrinos are calculated both analytically and
numerically, by keeping their leading (i.e., one-loop) RGE-induced effects.

\section{Two-loop RGE-induced corrections}

\begin{figure}[t]
  \centering
  \includegraphics[width=0.4\linewidth]{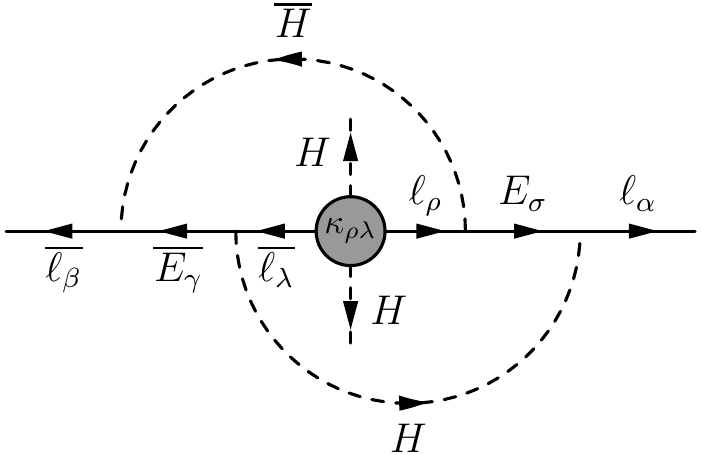}
  \caption{The dominant two-loop Feynman diagram that can increase the rank of
  the effective Majorana neutrino coupling matrix $\kappa$ from two to three
  because of the SM-like leptonic Yukawa interactions as described by Eq.~(1).}
  \label{Higgs-M}
\end{figure}

Given the SM-like Yukawa interactions in Eq.~(1) and the dimension-five Weinberg
operator as the origin of tiny neutrino masses in Eq.~(2), an exactly massless
neutrino running from a superhigh energy scale $\Lambda$ down to the Fermi scale
$\Lambda^{}_{\rm F}$ will stay massless provided only the one-loop RGE of the
effective Majorana neutrino coupling matrix $\kappa$ is
taken into account. The reason is simply that $m^{}_1 =0$ (or $m^{}_3 =0$) requires
the rank of $\kappa$ to be two, but the one-loop quantum corrections
to $\kappa$ do not change its rank. When the two-loop radiative corrections to
$\kappa$ are taken into consideration, however, Davidson {\it et al} have pointed
out that a {\it nontrivial} quantum effect described by the Feynman
diagram in Fig.~\ref{Higgs-M} can increase the rank of $\kappa$ from two to three,
and the contributions from all the other two-loop Feynman diagrams are qualitatively
trivial and thus quantitatively negligible \cite{Davidson:2006tg}. This interesting
observation has been confirmed by our recalculations along the same line of
thought. As a straightforward consequence, the initially vanishing neutrino mass
at $\Lambda$ will become nonzero at an energy scale below $\Lambda$ (e.g., at
the Fermi scale $\Lambda^{}_{\rm F}$) thanks to the two-loop RGE evolution.

\begin{figure}[t]
  \centering
  \includegraphics[width=0.9\linewidth]{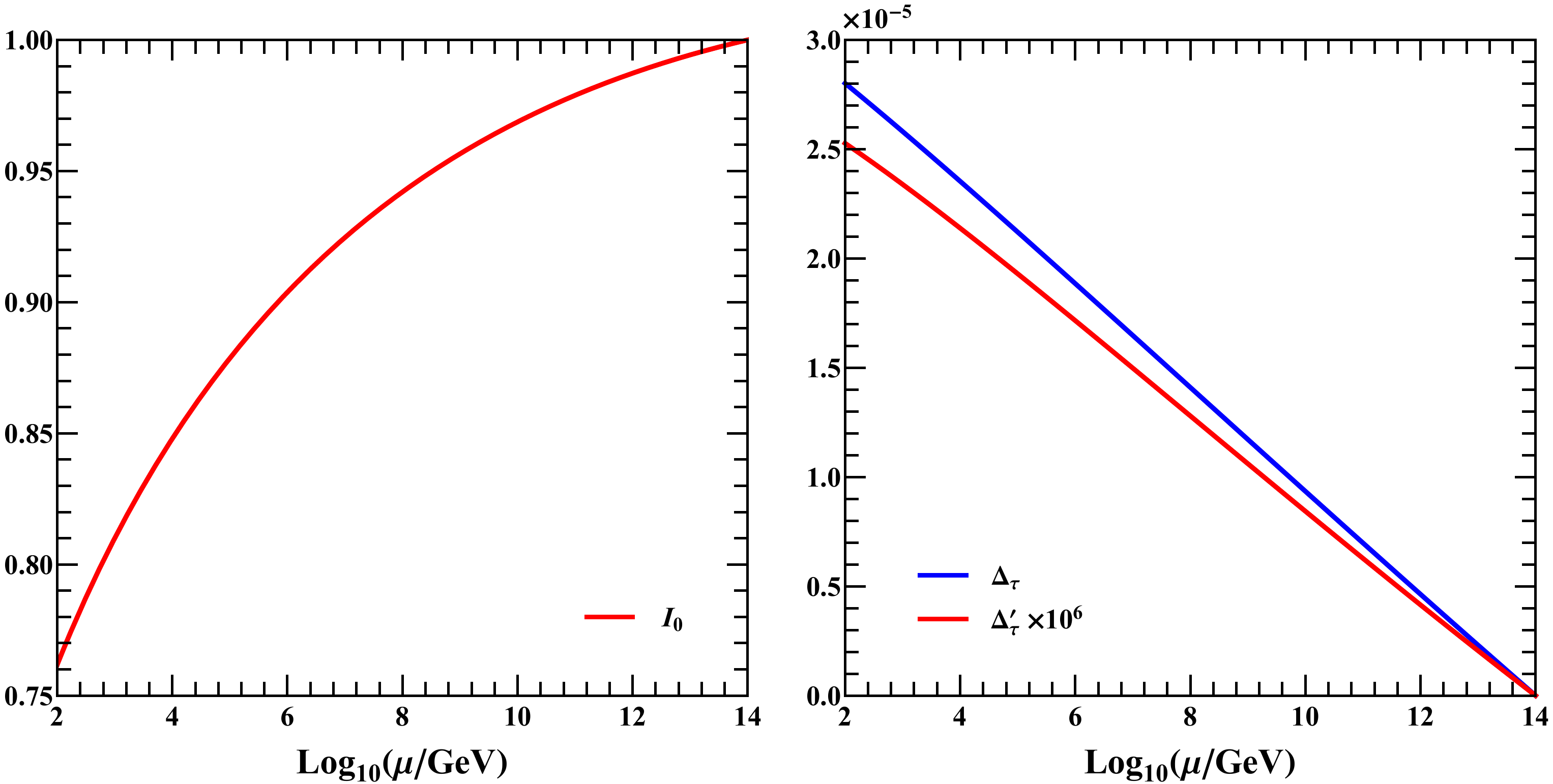}
  \caption{Changes of $I^{}_0$, $\Delta^{}_\tau$ and $\Delta^\prime_\tau$
  with the energy scale $\mu$ below $\Lambda \simeq 10^{14}$ GeV in the SM.}
  \label{loop function}
\end{figure}

To be explicit, we write out the RGE of $\kappa$ which includes both the one-loop
contributions and the nontrivial two-loop effect originating from Fig.~\ref{Higgs-M}
\cite{Davidson:2006tg}:
\begin{eqnarray}
16\pi^2 \frac{{\rm d}\kappa}{{\rm d}t} = \alpha^{}_\kappa \kappa - \frac{3}{2}
\left[ \left( Y^{}_l Y^\dagger_l \right) \kappa + \kappa \left( Y^{}_l Y^\dagger_l
\right)^T  \right] + \frac{1}{8\pi^2} \left( Y^{}_l Y^\dagger_l \right) \kappa
\left( Y^{}_l Y^\dagger_l \right)^T \; ,
\end{eqnarray}
where $t \equiv \ln{\left( \mu/\Lambda^{}_{\rm F} \right)}$ with $\mu$ being an
arbitrary renormalization scale between $\Lambda^{}_{\rm F}$ and $\Lambda$, and
$\alpha^{}_\kappa \simeq -3 g^2_2 + 6 y^2_t + \lambda$ with $g^{}_2$, $y^{}_t$
and $\lambda$ standing respectively for the $\rm SU(2)^{}_{\rm L}$ gauge coupling,
the top-quark Yukawa coupling and the Higges self-coupling constant.
It is obvious that the first two terms on the right-hand side of Eq.~(4) are the
one-loop contributions \cite{Chankowski:1993tx,Babu:1993qv,Antusch:2001ck,
Antusch:2005gp,Mei:2005qp,Ohlsson:2013xva}, and the last term is the
nontrivial two-loop contribution induced by Fig.~\ref{Higgs-M}. Without loss of
generality, we study the evolution of $M^{}_\nu = - \kappa \langle H \rangle^2$
from $\Lambda$ to $\Lambda^{}_{\rm F}$ in the basis where $Y^{}_l$ is taken to
be diagonal (i.e., $Y^{}_l = {\rm Diag}\{ y^{}_e, y^{}_\mu, y^{}_\tau \}$).
Since $Y^{}_l$ keeps diagonal during the RGE evolution \cite{Xing:2011zza}, we
integrate Eq.~(4) and arrive at
\begin{eqnarray}
M^{}_\nu (\Lambda^{}_{\rm F})  = I^{}_0 T^{}_l \left[ M^{}_\nu (\Lambda)
\circ \Omega \right] T^{}_l \;,
\end{eqnarray}
where $M^{}_\nu (\Lambda)$ and $M^{}_\nu (\Lambda^{}_{\rm F})$
stand respectively for the effective Majorana neutrino mass matrices at $\Lambda$
and $\Lambda^{}_{\rm F}$, the mathematical symbol ``$\circ$" denotes the
so-called Hadamard product (also known as the Schur product \cite{Horn2012})
which produces a new matrix by multiplying the elements in the same position
of the two original matrices with the same dimension [i.e., $\left( M^{}_\nu
\circ \Omega \right)^{}_{\alpha\beta} = \left( M^{}_\nu \right)^{}_{\alpha\beta}
\Omega^{}_{\alpha\beta} $], $T^{}_l = {\rm Diag} \{ I^{}_e, I^{}_\mu, I^{}_\tau \}$
is diagonal but flavor-dependent, and the loop functions $I^{}_0$, $I^{}_\alpha$
and $\Omega^{}_{\alpha\beta}$ (for $\alpha,\beta = e, \mu, \tau $) are defined as
\begin{eqnarray}
I^{}_0 \hspace{-0.2cm}&=&\hspace{-0.2cm} \exp \left[ - \frac{1}{16\pi^2}
\int^{\ln \left( \Lambda/\Lambda^{}_{\rm F} \right)}_{0} \alpha^{}_\kappa
\left(t\right) {\rm d}t \right] \;,
\nonumber \\
I^{}_\alpha \hspace{-0.2cm}&=&\hspace{-0.2cm} \exp \left[ \frac{3}{32\pi^2}
\int^{\ln \left( \Lambda/\Lambda^{}_{\rm F} \right)}_{0} y^2_\alpha
\left(t\right) {\rm d}t \right]\;,
\nonumber \\
\Omega^{}_{\alpha\beta} \hspace{-0.2cm}&=&\hspace{-0.2cm} \exp \left[
-\frac{1}{128\pi^4} \int^{\ln \left( \Lambda/\Lambda^{}_{\rm F} \right)}_{0}
y^2_\alpha \left(t\right) \hspace{0.05cm} y^2_\beta \left(t\right) {\rm d}t \right] \;.
\end{eqnarray}
It is clear that the one-loop effects described by $I^{}_0$ and $T^{}_l$ cannot
change the rank of $M^{}_\nu (\Lambda)$, but the nontrivial two-loop effect hidden
in $\Omega$ is able to increase the rank of $M^{}_\nu (\Lambda)$ from two to three
because its contribution to $M^{}_\nu (\Lambda)$ is not flavor-diagonal.
Given $y^2_e \ll y^2_\mu \ll y^2_\tau \ll 1$ in the SM~\cite{Xing:2019vks}, it
is very safe to make the $\tau$-dominance approximations as follows:
\begin{eqnarray}
T^{}_l \hspace{-0.2cm} &\simeq& \hspace{-0.2cm}
\left(\begin{matrix} 1 & 0 & 0 \cr 0 & 1 & 0 \cr 0 & 0 &
1 + \Delta^{}_\tau
\end{matrix}\right) \; ,
\nonumber \\
\Omega \hspace{-0.2cm} &\simeq& \hspace{-0.2cm}
\left(\begin{matrix} 1 & 1 & 1 \cr
1 & 1 & 1 \cr 1 & 1 & 1 - \Delta^\prime_\tau \end{matrix}\right) \;,
\end{eqnarray}
where
\begin{eqnarray}
\Delta^{}_\tau \hspace{-0.2cm}&=&\hspace{-0.2cm} \frac{3}{32\pi^2}
\int^{\ln \left( \Lambda/\Lambda^{}_{\rm F} \right)}_{0} y^2_\tau
\left(t\right) {\rm d}t \;,
\nonumber \\
\Delta^\prime_\tau \hspace{-0.2cm}&=&\hspace{-0.2cm} \frac{1}{128\pi^4}
\int^{\ln \left( \Lambda/\Lambda^{}_{\rm F} \right)}_{0} y^4_\tau
\left(t\right)  {\rm d}t \;.
\end{eqnarray}
So $\Delta^{}_\tau$ contributes to every element in the third row and the
third column of $M^{}_\nu (\Lambda)$, but $\Delta^\prime_\tau$ only affects
the (3,3) element of $M^{}_\nu (\Lambda)$. The values of $\Delta^{}_\tau$
and $\Delta^\prime_\tau$ are both positive in the SM, and their dependence
on the energy scale $\mu$ is shown in Fig.~\ref{loop function}, where the
dependence of $I^{}_0$ on $\mu$ is also illustrated. One can immediately see that
$\Delta^\prime_\tau$ is roughly $10^6$ times smaller than $\Delta^{}_\tau$;
and their magnitudes are of $\mathcal{O}\left(10^{-11}\right)$ and
$\mathcal{O}\left(10^{-5}\right)$, respectively, when $\Lambda \simeq
10^{14}$ GeV is fixed and $\mu \lesssim 10^{10}$ GeV holds.

In the chosen basis with $Y^{}_l$ being diagonal, the effective Majorana
neutrino mass matrix $M^{}_\nu$ can be reconstructed in terms of the PMNS
matrix $U$ and the diagonal neutrino mass matrix
$D^{}_\nu = {\rm Diag}\{m^{}_1, m^{}_2, m^{}_3\}$ at a given energy
scale $\Lambda$. Namely, $M^{}_\nu = U D^{}_\nu U^T$. Substituting both
Eq.~(7) and the decompositions of $M^{}_\nu$ at $\Lambda$ and
$\Lambda^{}_{\rm F}$ into Eq.~(5), we obtain the relationship
\begin{eqnarray}
D^{}_\nu \left( \Lambda^{}_{\rm F} \right) \simeq I^{}_0
U^\dagger \left( \Lambda^{}_{\rm F} \right) \left[ U D^{}_\nu U^{T} + \Delta^{}_\tau \left(\begin{matrix} 0 & 0 &
\sum\limits^{}_i m^{}_i U^{}_{e i} U^{}_{\tau i} \cr 0 & 0 &
\sum\limits^{}_i m^{}_i U^{}_{\mu i} U^{}_{\tau i} \cr
\sum\limits^{}_i m^{}_i U^{}_{e i} U^{}_{\tau i} &
\sum\limits^{}_i m^{}_i U^{}_{\mu i} U^{}_{\tau i} & \left(
2 - r^{}_\tau \right) \sum\limits^{}_i m^{}_i
U^{2}_{\tau i} \end{matrix}\right) \right]_{\Lambda} U^{*} \left( \Lambda^{}_{\rm F} \right)\;,
\end{eqnarray}
where $r^{}_\tau \equiv \Delta^\prime_\tau/\Delta^{}_\tau$ signifies the
tiny two-loop RGE-induced effect. If one of the three neutrinos is exactly
massless at $\Lambda$, Eq.~(9) tells us that the determinant of
$M^{}_\nu (\Lambda^{}_{\rm F}) = \left( U D^{}_\nu U^T
\right)^{}_{\Lambda^{}_{\rm F}}$ must be proportional to $\Delta^{\prime}_\tau$.
It is therefore the diagonal part of Eq.~(9) that allows us to calculate a
nonzero result of $m^{}_1$ (or $m^{}_3$) and the corresponding Majorana CP phase
at $\Lambda^{}_{\rm F}$ from $m^{}_1 =0$ (or $m^{}_3 =0$) at $\Lambda$. In the
leading-order approximation, we arrive at
\begin{eqnarray}
m^{}_1 \simeq
-\Delta^\prime_\tau \left[ m^{}_2 \left(U^{}_{\tau 2} U^{*}_{\tau 1}\right)^2
+ m^{}_3 \left(U^{}_{\tau 3} U^{*}_{\tau 1}\right)^2 \right] \;
\end{eqnarray}
in the normal neutrino mass ordering case with $m^{}_1 (\Lambda) =0$; or
\begin{eqnarray}
m^{}_3 \simeq
-\Delta^\prime_\tau \left[ m^{}_1 \left(U^{}_{\tau 1} U^{*}_{\tau 3}\right)^2
+ m^{}_2 \left(U^{}_{\tau 2} U^{*}_{\tau 3}\right)^2 \right] \;
\end{eqnarray}
in the inverted neutrino mass ordering case with $m^{}_3 (\Lambda) =0$, where
all the neutrino masses and flavor mixing parameters are
defined at the Fermi scale $\Lambda^{}_{\rm F}$. In view of the fact
that $m^{}_i$ (for $i = 1,2,3$) must be real and positive, one may determine
the Majorana CP phase associated with $m^{}_1$ (or $m^{}_3$)
at $\Lambda^{}_{\rm F}$ by taking the imaginary part of Eq.~(10) or
Eq.~(11) to be vanishing, and then obtain the explicit expression of $m^{}_1$ (or
$m^{}_3$) from the real part of Eq.~(10) or Eq.~(11).

Since the Majorana CP phases of the $3\times 3$ PMNS matrix $U$ at a given
superhigh energy scale $\Lambda$ depend on its phase convention,
let us take the following popular parametrization \cite{Tanabashi:2018oca}:
\begin{eqnarray}
U = P^{}_l \left(\begin{matrix} c^{}_{12}c^{}_{13} & s^{}_{12}c^{}_{13} &
s^{}_{13}e^{-{\rm i}\delta} \cr -s^{}_{12}c^{}_{23} - c^{}_{12}s^{}_{13}
s^{}_{23}e^{{\rm i}\delta} & c^{}_{12}c^{}_{23} - s^{}_{12}s^{}_{13}s^{}_{23}
e^{{\rm i}\delta} & c^{}_{13}s^{}_{23} \cr s^{}_{12}s^{}_{23} - c^{}_{12}
s^{}_{13}c^{}_{23}e^{{\rm i}\delta} & -c^{}_{12}s^{}_{23} - s^{}_{12}s^{}_{13}
c^{}_{23}e^{{i}\delta} & c^{}_{13}c^{}_{23} \end{matrix}\right) P^{}_\nu \;,
\end{eqnarray}
in which $c^{}_{ij} \equiv \cos \theta^{}_{ij}$ and $s^{}_{ij} \equiv \sin
\theta^{}_{ij}$ (for $ij=12,13,23$) with $\theta^{}_{ij}$ lying in the first
quadrant, $\delta$ is the so-called Dirac CP phase, $P^{}_l \equiv {\rm Diag}
\{ e^{{\rm i}\phi^{}_e}, e^{{\rm i}\phi^{}_\mu}, e^{{\rm i}\phi^{}_\tau} \}$
with $\phi^{}_e$, $\phi^{}_\mu$ and $\phi^{}_\tau$ being the unphysical phases
associated with the charged-lepton fields, and $P^{}_\nu$ is a
phase matrix containing two independent Majorana CP phases. Here we choose
the phase convention of $P^{}_\nu$ as
\begin{eqnarray}
P^{}_\nu \equiv \left\{ \begin{matrix}
{\rm Diag} \{ e^{{\rm i}\rho}, e^{{\rm i}\sigma}, 1 \} \; , \quad
\left(m^{}_1 < m^{}_2 < m^{}_3\right) \; , \cr
{\rm Diag} \{ 1, e^{{\rm i}\sigma}, e^{{\rm i}\varrho} \} \; , \quad
\left(m^{}_3 < m^{}_1 < m^{}_2\right) \; , \cr \end{matrix}
\right.
\end{eqnarray}
corresponding to the normal and inverted neutrino mass ordering cases, respectively.
Since $\rho$ (or $\varrho$) can always be removed in the $m^{}_1 =0$ (or
$m^{}_3 =0$) limit, only a single Majorana CP phase $\sigma$ survives when
$M^{}_\nu$ is a rank-two mass matrix. At the Fermi scale $\Lambda^{}_{\rm F}$,
the PMNS matrix $U (\Lambda^{}_{\rm F})$ can be parametrized in the same form
as that of $U (\Lambda)$. It is convenient to define
\begin{eqnarray}
&& \Delta \theta^{}_{ij} \equiv \theta^{}_{ij} (\Lambda^{}_{\rm F}) -
\theta^{}_{ij} (\Lambda) \; , \quad
\Delta \delta \equiv \delta (\Lambda^{}_{\rm F}) - \delta (\Lambda) \; , \quad
\Delta \sigma \equiv \sigma (\Lambda^{}_{\rm F}) - \sigma (\Lambda) \; ,
\hspace{0.8cm}
\nonumber \\
&& \Delta \phi^{}_\alpha \equiv \phi^{}_\alpha (\Lambda^{}_{\rm F}) -
\phi^{}_\alpha (\Lambda) \; , \quad
\Delta \rho \equiv \rho (\Lambda^{}_{\rm F}) - \rho (\Lambda) \; , \quad
\Delta \varrho \equiv \varrho (\Lambda^{}_{\rm F}) - \varrho (\Lambda) \; ,
\end{eqnarray}
so as to describe the strengths of the RGE-induced corrections to the relevant
flavor mixing angles and phase parameters. The smallness of such quantum
corrections, which are expected to be proportional to either $\Delta^{}_\tau$ or
$\Delta^\prime_\tau$, makes it reasonable to treat them as small perturbations
in the leading-order analytical approximations.

\begin{center}
{\bf (A) The $m^{}_1 = 0$ case at $\Lambda$}
\end{center}

We first calculate the finite values of $m^{}_1$ and $\rho$ at $\Lambda^{}_{\rm F}$
which originate from $m^{}_1 = 0$ at $\Lambda$ via the two-loop RGE-induced
effect. Substituting Eqs.~(12) and (13) into Eq.~(10), we obtain the following
results after a lengthy but straightforward calculation:
\begin{eqnarray}
m^{}_1 \hspace{-0.2cm}&\simeq&\hspace{-0.2cm} \Delta^\prime_\tau \left(
\sin^2{\theta^{}_{12}} \sin^2{\theta^{}_{23}} + \cos^2{\theta^{}_{12}}
\sin^2{\theta^{}_{13}} \cos^2{\theta^{}_{23}} - \frac{1}{2} \sin{2\theta^{}_{12}}
\sin{\theta^{}_{13}} \sin{2\theta^{}_{23}} \cos{\delta} \right)
\sqrt{ \mathcal{F}^{}_1 } \;, \hspace{0.2cm}
\nonumber \\
2\rho \hspace{-0.2cm}&\simeq&\hspace{-0.2cm} \arctan{\left(
\frac{\mathcal{A}^{}_1}{\mathcal{B}^{}_1} \right)} \;,
\end{eqnarray}
where
\begin{eqnarray}
\mathcal{F}^{}_1 \hspace{-0.2cm}&=&\hspace{-0.2cm} m^2_2 \left( \cos^2{\theta^{}_{12}}
\sin^2{\theta^{}_{23}} + \sin^2{\theta^{}_{12}} \sin^2{\theta^{}_{13}}
\cos^2{\theta^{}_{23}} + \frac{1}{2} \sin{2\theta^{}_{12}} \sin{\theta^{}_{13}}
\sin{2\theta^{}_{23}} \cos{\delta} \right)^2
\nonumber
\\
\hspace{-0.2cm}&&\hspace{-0.2cm} + m^2_3 \cos^4{\theta^{}_{13}}
\cos^4{\theta^{}_{23}} + 2 m^{}_2 m^{}_3 \cos^2{\theta^{}_{13}}
\cos^2{\theta^{}_{23}} \left[ \vphantom{\frac{1}{1}} \sin^2{\theta^{}_{12}}
\sin^2{\theta^{}_{13}}\cos^2{\theta^{}_{23}} \cos{2\left( \sigma + \delta \right)}
\right. \hspace{0.4cm}
\nonumber
\\
\hspace{-0.2cm}&&\hspace{-0.2cm} + \left. \cos^2{\theta^{}_{12}}
\sin^2{\theta^{}_{23}} \cos{2\sigma} + \frac{1}{2} \sin{2\theta^{}_{12}}
\sin{\theta^{}_{13}} \sin{2\theta^{}_{23}} \cos{\left( 2\sigma + \delta \right) }
\right] \;,
\end{eqnarray}
and
\begin{eqnarray}
\mathcal{A}^{}_1 \hspace{-0.2cm}&=&\hspace{-0.2cm} - m^{}_3 \sin{2\theta^{}_{13}}
\cos^2{\theta^{}_{23}} \left( \sin{2\theta^{}_{12}} \cos{\theta^{}_{13}}
\sin{2\theta^{}_{23}} \sin{\delta} - \cos^2{\theta^{}_{12}} \sin{2\theta^{}_{13}}
\cos^2{\theta^{}_{23}} \sin{2\delta} \right)
\nonumber
\\
\hspace{-0.2cm}&&\hspace{-0.2cm} - m^{}_2 \left\{ 2 \sin{2\theta^{}_{12}}
\sin{\theta^{}_{13}} \sin{2\theta^{}_{23}} \left( \sin^2{\theta^{}_{13}}
\cos^2{\theta^{}_{23}} -\sin^2{\theta^{}_{23}} \right) \left[ \cos^2{\theta^{}_{12}}
\sin{\left( 2\sigma - \delta \right)} \right.\right.
\nonumber
\\
\hspace{-0.2cm}&&\hspace{-0.2cm} - \left. \sin^2{\theta^{}_{12}} \sin{\left(
2\sigma + \delta \right)} \right] + \sin^2{2\theta^{}_{12}} \sin{2\sigma}
\left( \sin^4{\theta^{}_{23}} + \sin^4{\theta^{}_{13}} \cos^4{\theta^{}_{23}} -
\sin^2{\theta^{}_{13}} \sin^2{2\theta^{}_{23}} \right)
\nonumber
\\
\hspace{-0.2cm}&&\hspace{-0.2cm} + \left. \sin^2{\theta^{}_{13}}
\sin^2{2\theta^{}_{23}} \left[ \cos^4{\theta^{}_{12}} \sin{2\left(
\sigma -\delta \right)} + \sin^4{\theta^{}_{12}} \sin{2\left(
\sigma + \delta \right)} \right] \right\} \;,
\nonumber
\\
\mathcal{B}^{}_1 \hspace{-0.2cm}&=&\hspace{-0.2cm} - m^{}_3 \left[ \sin^2{\theta^{}_{12}}
\cos^2{\theta^{}_{13}} \sin^2{2\theta^{}_{23}} - \sin{2\theta^{}_{13}}
\cos^2{\theta^{}_{23}} \left( \sin{2\theta^{}_{12}} \cos{\theta^{}_{13}}
\sin{2\theta^{}_{23}} \cos{\delta}  \right.\right.
\nonumber
\\
\hspace{-0.2cm}&&\hspace{-0.2cm} - \left.\left. \cos^2{\theta^{}_{12}}
\sin{2\theta^{}_{13}} \cos^2{\theta^{}_{23}} \cos{2\delta} \right) \right] -
m^{}_2 \left\{ 2 \sin{2\theta^{}_{12}} \sin{\theta^{}_{13}} \sin{2\theta^{}_{23}}
\left( \sin^2{\theta^{}_{13}} \cos^2{\theta^{}_{23}} \right.\right.
\nonumber
\\
\hspace{-0.2cm}&&\hspace{-0.2cm} - \left. \sin^2{\theta^{}_{23}} \right)
\left[ \cos^2{\theta^{}_{12}} \cos{\left( 2\sigma - \delta \right)} -
\sin^2{\theta^{}_{12}} \cos{\left( 2\sigma + \delta \right)} \right] +
\sin^2{2\theta^{}_{12}} \cos{2\sigma}
\nonumber
\\
\hspace{-0.2cm}&&\hspace{-0.2cm} \times \left( \sin^4{\theta^{}_{23}} +
\sin^4{\theta^{}_{13}} \cos^4{\theta^{}_{23}} - \sin^2{\theta^{}_{13}}
\sin^2{2\theta^{}_{23}} \right) + \sin^2{\theta^{}_{13}} \sin^2{2\theta^{}_{23}}
\nonumber
\\
\hspace{-0.2cm}&&\hspace{-0.2cm} \times \left. \left[ \cos^4{\theta^{}_{12}}
\cos{2\left( \sigma -\delta \right)} + \sin^4{\theta^{}_{12}} \cos{2\left(
\sigma + \delta \right)} \right] \right\} \;.
\end{eqnarray}

\begin{center}
{\bf (B) The $m^{}_3 = 0$ case at $\Lambda$}
\end{center}

In the inverted neutrino mass ordering case with $m^{}_3 =0$ at $\Lambda$,
the finite results of $m^{}_3$ and $\varrho$ at $\Lambda^{}_{\rm F}$ are
similarly obtained as follows:
\begin{eqnarray}
m^{}_3 \hspace{-0.2cm}&\simeq&\hspace{-0.2cm} \Delta^\prime_\tau
\left(\cos^2{\theta^{}_{13}} \cos^2{\theta^{}_{23}} \right)
\sqrt{ \mathcal{F}^{}_3 } \;,
\nonumber
\\
2\varrho \hspace{-0.2cm}&\simeq&\hspace{-0.2cm} \arctan{\left(
\frac{\mathcal{A}^{}_3}{\mathcal{B}^{}_3} \right)} \;,
\end{eqnarray}
where
\begin{eqnarray}
\mathcal{F}^{}_3 \hspace{-0.2cm}&=&\hspace{-0.2cm} m^2_1 \left( \sin^2{\theta^{}_{12}}
\sin^2{\theta^{}_{23}} + \cos^2{\theta^{}_{12}} \sin^2{\theta^{}_{13}}
\cos^2{\theta^{}_{23}} - \frac{1}{2} \sin{2\theta^{}_{12}} \sin{\theta^{}_{13}}
\sin{2\theta^{}_{23}} \cos{\delta} \right)^2
\nonumber
\\
\hspace{-0.2cm}&&\hspace{-0.2cm} + m^2_2 \left( \cos^2{\theta^{}_{12}}
\sin^2{\theta^{}_{23}} + \sin^2{\theta^{}_{12}} \sin^2{\theta^{}_{13}}
\cos^2{\theta^{}_{23}} + \frac{1}{2} \sin{2\theta^{}_{12}} \sin{\theta^{}_{13}}
\sin{2\theta^{}_{23}} \cos{\delta} \right)^2
\nonumber
\\
\hspace{-0.2cm}&&\hspace{-0.2cm} + \frac{1}{2} m^{}_1 m^{}_2 \left\{
2\sin{2\theta^{}_{12}} \sin{\theta^{}_{13}} \sin{2\theta^{}_{23}} \left(
\sin^2{\theta^{}_{23}} - \sin^2{\theta^{}_{13}} \cos^2{\theta^{}_{23}} \right)
\left[ \sin^2{\theta^{}_{12}} \cos{\left( 2\sigma + \delta \right)} \right.\right.
\nonumber
\\
\hspace{-0.2cm}&&\hspace{-0.2cm} - \left. \cos^2{\theta^{}_{12}} \cos{\left(
2\sigma -\delta \right)} \right] + \sin^2{2\theta^{}_{12}} \cos{2\sigma} \left(
\sin^4{\theta^{}_{23}} + \sin^4{\theta^{}_{13}} \cos^4{\theta^{}_{23}} -
\sin^2{\theta^{}_{13}} \sin^2{2\theta^{}_{23}} \right) \hspace{0.2cm}
\nonumber
\\
\hspace{-0.2cm}&&\hspace{-0.2cm} + \left. \sin^2{\theta^{}_{13}}
\sin^2{2\theta^{}_{23}} \left[ \sin^4{\theta^{}_{12}} \cos{2\left( \sigma +
\delta \right)} + \cos^4{\theta^{}_{12}} \cos{2\left( \sigma - \delta \right)}
\right] \right\} \;,
\end{eqnarray}
and
\begin{eqnarray}
\mathcal{A}^{}_3 \hspace{-0.2cm}&=&\hspace{-0.2cm} - m^{}_1 \left( 2
\cos^2{\theta^{}_{12}} \sin^2{\theta^{}_{13}} \cos^2{\theta^{}_{23}} \sin{2\delta}
- \sin{2\theta^{}_{12}} \sin{\theta^{}_{13}} \sin{2\theta^{}_{23}} \sin{\delta}
\right)
\nonumber
\\
\hspace{-0.2cm}&&\hspace{-0.2cm} - m^{}_2 \left[ 2\cos^2{\theta^{}_{12}}
\sin^2{\theta^{}_{23}} \sin{2\sigma} + 2\sin^2{\theta^{}_{12}}
\sin^2{\theta^{}_{13}} \cos^2{\theta^{}_{23}} \sin{2\left( \sigma + \delta
\right)} \right.
\nonumber
\\
\hspace{-0.2cm}&&\hspace{-0.2cm} + \left. \sin{2\theta^{}_{12}}
\sin{\theta^{}_{13}} \sin{2\theta^{}_{23}} \sin{\left( 2\sigma + \delta
\right)} \right] \; ,
\nonumber
\\
\mathcal{B}^{}_3 \hspace{-0.2cm}&=&\hspace{-0.2cm} - m^{}_1 \left( 2
\sin^2{\theta^{}_{12}} \sin^2{\theta^{}_{23}} + 2 \cos^2{\theta^{}_{12}}
\sin^2{\theta^{}_{13}} \cos^2{\theta^{}_{23}} \cos{2\delta} - \sin{2\theta^{}_{12}}
\sin{\theta^{}_{13}} \sin{2\theta^{}_{23}} \cos{\delta}  \right) \hspace{0.2cm}
\nonumber
\\
\hspace{-0.2cm}&&\hspace{-0.2cm} - m^{}_2 \left[ 2\cos^2{\theta^{}_{12}}
\sin^2{\theta^{}_{23}} \cos{2\sigma} + 2\sin^2{\theta^{}_{12}} \sin^2{\theta^{}_{13}}
\cos^2{\theta^{}_{23}} \cos{2\left( \sigma + \delta \right)} \right.
\nonumber
\\
\hspace{-0.2cm}&&\hspace{-0.2cm} + \left. \sin{2\theta^{}_{12}} \sin{\theta^{}_{13}}
\sin{2\theta^{}_{23}} \cos{\left( 2\sigma + \delta \right)} \right] \; .
\end{eqnarray}
We remark that all the neutrino masses and flavor mixing parameters appearing
in Eqs.~(15)---(20) take their values at the Fermi scale $\Lambda^{}_{\rm F}$.
Such a treatment is advantageous to our numerical estimates because it allows us
to figure out the radiatively generated values of $m^{}_1$ and $\rho$ (or $m^{}_3$
and $\varrho$) at $\Lambda^{}_{\rm F}$ by directly inputting the experimental
data at low energies. Different from $m^{}_1$ (or $m^{}_3$), whose running effect from
$\Lambda$ to $\Lambda^{}_{\rm F}$ is apparently measured by the value
of $\Delta^\prime_\tau$, the Majorana CP phase $\rho$ (or $\varrho$) is essentially
insensitive to a change of the energy scale. This phase parameter is
not well defined when $m^{}_1 =0$ (or $m^{}_3 =0$) exactly holds at $\Lambda$,
but it will become physical soon after the vanishing neutrino mass acquires a
tiny nonzero value just a bit below $\Lambda$. Once $\rho$ (or $\varrho$) is radiatively
generated together with $m^{}_1$ (or $m^{}_3$), it will almost keep unchanged until
$\Lambda^{}_{\rm F}$.

At this point it is also worth remarking that our analytical results in
Eqs.~(15)---(20) are essentially new. In comparison, Davidson {\it et al} have
only presented the considerably simplified expression of $m^{}_1 e^{{\rm i} 2\rho}$
(or $m^{}_3 e^{{\rm i} 2\varrho}$) by explicitly taking $\sin\theta^{}_{12} =
1/\sqrt{3}$, $\sin\theta^{}_{13} \ll 1$ and $\sin\theta^{}_{23} = 1/\sqrt{2}$ in
Ref.~\cite{Davidson:2006tg} to give the reader a ball park feeling
of the two-loop RGE-induced effect. The latest global analysis of currently available
neutrino oscillation data \cite{Capozzi:2020qhw}, in which the T2K collaboration's
$3\sigma$ evidence for $\delta \neq 0$ (or $\pi$) \cite{Abe:2019vii} has been included,
yields the best-fit values
\begin{eqnarray}
\sin^2 \theta^{}_{12} = \left\{ \begin{array}{l} 0.305 \\ 0.303 \end{array} \right.
\;, \quad \sin^2 \theta^{}_{13} = \left\{ \begin{array}{l} 0.0222 \\ 0.0223
\end{array} \right. \;, \quad \sin^2 \theta^{}_{23} = \left\{ \begin{array}{l} 0.545
\\ 0.551 \end{array} \right. \;,\quad \delta = \left\{ \begin{array}{l} 1.28\pi
\\ 1.52\pi \end{array} \right. \;,
\end{eqnarray}
and
\begin{eqnarray}
\delta m^2 = \left\{ \begin{array}{l} 7.34\times 10^{-5} ~{\rm eV^2} \\
7.34\times 10^{-5} ~{\rm eV^2} \end{array} \right. \;, \quad
\Delta m^2 = \left\{ \begin{array}{l} + 2.485 \times 10^{-3} ~{\rm eV^2} \\
-2.465 \times 10^{-3} ~{\rm eV^2} \end{array} \right. \;,
\end{eqnarray}
where both the normal neutrino mass ordering (upper values) and the
inverted one (lower values) have been taken into account, and the two
neutrino mass-squared differences are defined as $\delta m^2 \equiv
m^2_2 - m^2_1$ and $\Delta m^2 \equiv m^2_3 - \left(m^2_1 + m^2_2\right)/2$.
These results will be used in our subsequent numerical estimates of
$m^{}_1$ and $\rho$ (or $m^{}_3$ and $\varrho$) at $\Lambda^{}_{\rm F}$,
which are generated from $m^{}_1 =0$ (or $m^{}_3 =0$) at
$\Lambda \gg \Lambda^{}_{\rm F}$ via the two-loop RGE evolution.
\begin{figure}[t]
  \centering
  \includegraphics[width=0.98\linewidth]{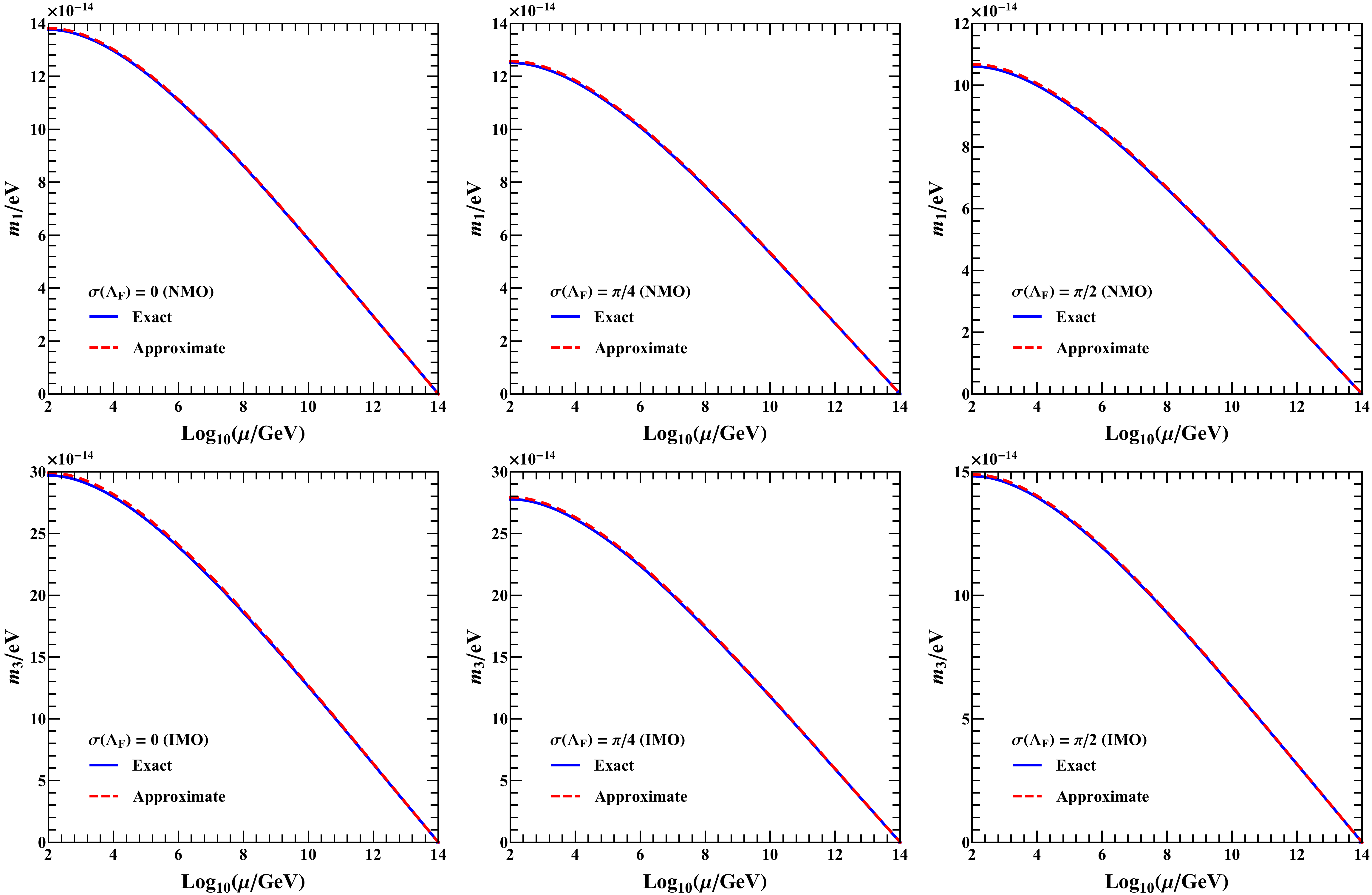}
  \caption{The numerical result of $m^{}_1$ (or $m^{}_3$) at a given energy scale
  above $\Lambda^{}_{\rm F}$, which is radiatively generated from $m^{}_1 = 0$ (or
  $m^{}_3 = 0$) at $\Lambda \simeq 10^{14}$ GeV in the normal (or inverted) neutrino
  mass ordering case with $\sigma (\Lambda^{}_{\rm F}) = 0$, $\pi/4$ or $\pi/2$.}
  \label{mlightest}
\end{figure}

\begin{figure}[t]
  \centering
  \includegraphics[width=0.945\linewidth]{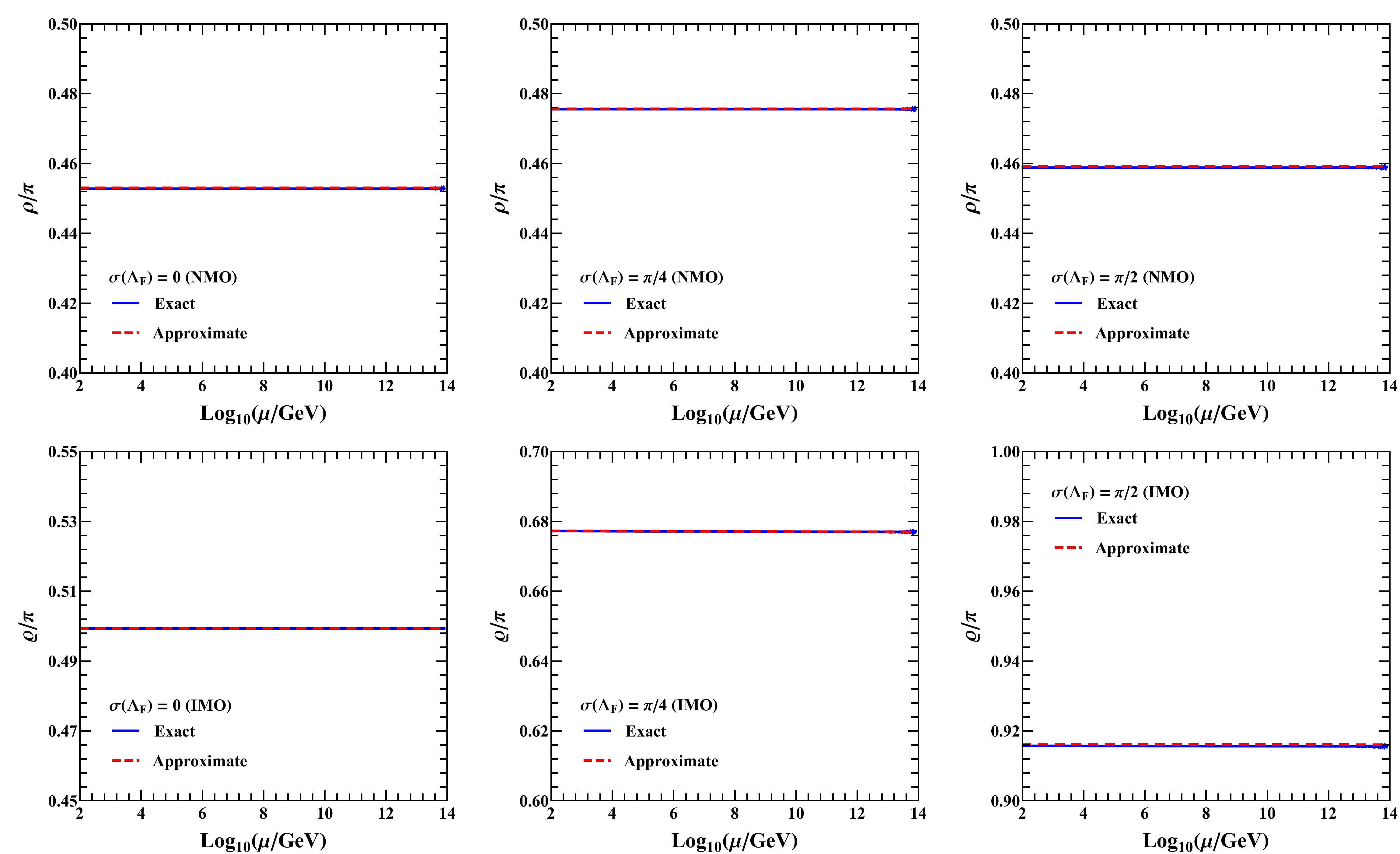}
  \caption{The numerical result of $\rho$ (or $\varrho$) at a given energy scale
  above $\Lambda^{}_{\rm F}$, which is radiatively generated from $m^{}_1 = 0$ (or
  $m^{}_3 = 0$) at $\Lambda \simeq 10^{14}$ GeV in the normal (or inverted) neutrino
  mass ordering case with $\sigma (\Lambda^{}_{\rm F}) = 0$, $\pi/4$ or $\pi/2$.}
  \label{rho}
\end{figure}

To compute the evolution of $m^{}_1$ and $\rho$ (or $m^{}_3$ and $\varrho$)
with the energy scale $\mu$, we incorporate the two-loop RGE of $\kappa$ described
by Eq.~(4) into those already known two-loop RGEs of the gauge couplings, the
quark and charged-lepton Yukawa couplings and the Higgs self-coupling constant in
the SM \cite{Machacek:1983tz,Machacek:1983fi,Machacek:1984zw,Luo:2002ey}. Then with
$m^{}_1 =0$ (or $m^{}_3 =0$) being an input at $\Lambda$, one may choose the initial
values of all the other neutrino parameters at $\Lambda$ in such a way that the
best-fit values of $\theta^{}_{12}$, $\theta^{}_{13}$, $\theta^{}_{23}$, $\delta$,
$\delta m^2$ and $\Delta m^2$ shown in Eqs.~(21) and (22) can be achieved at
$\Lambda^{}_{\rm F}$, where the other Majorana CP phase $\sigma$ is required to
acquire a special value $0$, $\pi/4$ or $\pi/2$. The exact numerical results of
$m^{}_1$ and $\rho$ (or $m^{}_3$ and $\varrho$) in the normal (or inverted)
neutrino mass ordering case are obtained by numerically solving the full set of
two-loop RGEs, and they are explicitly plotted in Figs.~\ref{mlightest}
and \ref{rho}. To compare, the approximate numerical results based on our
analytical approximations in Eqs.~(15)---(20) are also illustrated in the
same figures. In addition, we list the results of $m^{}_1$ and $\rho$ (or $m^{}_3$
and $\varrho$) at $\Lambda^{}_{\rm F}$ in Table~\ref{tvl}, where
the values given in the parentheses are obtained by numerically solving the
two-loop RGEs.
\begin{table}[h!]
\centering
\caption{The values of $m^{}_1$ and $\rho$ (or $m^{}_3$ and $\varrho$) at
$\Lambda^{}_{\rm F} \simeq 10^2$ GeV, which are radiatively generated from
$m^{}_1 =0$ (or $m^{}_3 =0$) at $\Lambda \simeq 10^{14}$ GeV
in the normal (or inverted) neutrino mass ordering case with $\sigma (\Lambda^{}_{\rm F})
= 0$, $\pi/4$ or $\pi/2$. The corresponding results given in the parentheses are
obtained by numerically solving the two-loop RGEs.}
\vspace{0.3cm}
\begin{tabular}{c|cccc}
\hline\hline
 & $\sigma \left( \Lambda^{}_{\rm F} \right)/\pi$ & $0$ & $1/4$ & $1/2$ \\
\hline
\multirow{2}{*}{NMO} & $m^{}_1 \left( \Lambda^{}_{\rm F} \right)/10^{-13} {\rm eV}$
& 1.382 (1.377) & 1.258 (1.251) & 1.068 (1.061) \\
& $\rho \left( \Lambda^{}_{\rm F} \right)/\pi$ & 0.453 (0.453) & 0.476 (0.476) &
0.459 (0.459) \\
\hline
\multirow{2}{*}{IMO} & $m^{}_3 \left( \Lambda^{}_{\rm F} \right)/10^{-13} {\rm eV}$
& 2.991 (2.969) & 2.793 (2.777) & 1.489 (1.482) \\
& $\varrho \left( \Lambda^{}_{\rm F} \right)/\pi$ & 0.499 (0.499) & 0.677 (0.677)
& 0.916 (0.916) \\
\hline\hline
\end{tabular}
\label{tvl}
\end{table}

It is clear that our analytical approximations made in Eqs.~(15)---(20) are in good
agreement with the results obtained by numerically solving the two-loop RGEs, and
the relative accuracy is at the $\mathcal{O}$(1\textperthousand) level.
Fig.~\ref{mlightest} and Table~\ref{tvl} tell us that the value of $m^{}_1$
(or $m^{}_3$) at $\Lambda^{}_{\rm F}$ is about $10^{-13}$ eV, a result which
coincides with the previous estimate made in Ref.~\cite{Davidson:2006tg}.
From Fig.~\ref{rho} or Table~\ref{tvl}, one can see that $\rho$ (or $\varrho$)
has acquired a physical value at an energy scale just a bit below $\Lambda$,
and this value is essentially insensitive to the two-loop RGE evolution between
$\Lambda$ and $\Lambda^{}_{\rm F}$ in the SM. This interesting observation is
new, both analytically and numerically.
It is obvious that the input of the nontrivial Majorana CP phase $\sigma$ in the
$m^{}_1 =0$ (or $m^{}_3 =0$) limit at $\Lambda$ may quantitatively affect the radiative
generation of a nonzero value of $m^{}_1$ (or $m^{}_3$) and a physical value
of $\rho$ (or $\varrho$) at lower energies. That is why invoking a proper flavor
symmetry (e.g., the $\mu$-$\tau$ reflection symmetry \cite{Xing:2015fdg}) may help
to fix or constrain the value of $\sigma$ at $\Lambda$.

\section{Initially nonzero flavor parameters}

As a nontrivial by-product, the one-loop relations between those initially nonzero
flavor parameters at $\Lambda$ and their counterparts at $\Lambda^{}_{\rm F}$
will be established here in the case of either $m^{}_1 (\Lambda) = 0$
or $m^{}_3 (\Lambda) = 0$. It is unnecessary to consider the
two-loop RGE-induced effects on those parameters, simply because such effects
have no way to compete with the one-loop contributions. So we simply take
$r^{}_\tau = 0$ in Eq.~(9) to switch off very tiny contributions from the two-loop
term. The strategy of deriving the one-loop evolution of those initially nonzero 
neutrino masses, flavor mixing angles and CP-violating phases is as follows:
1) with the help of Eqs.~(12) and (13), we expand $U \left(\Lambda \right)$ in
Eq.~(9) in terms of the small quantities defined in Eq.~(14) by only keeping those 
leading-order terms; 2) then we obtain ten independent linear equations which 
contain ten parameters $m^{}_{2} \left( \Lambda^{}_{\rm EW} \right)$, $m^{}_{3}
\left( \Lambda^{}_{\rm EW} \right)$ [or $m^{}_{1} \left(
\Lambda^{}_{\rm EW} \right)$], $\Delta \theta^{}_{12}$, $\Delta \theta^{}_{13}$,
$\Delta \theta^{}_{23}$, $\Delta \delta$, $\Delta \sigma$, $\phi^{}_e$, $\phi^{}_\mu$
and $\phi^{}_\tau$ from the real and imaginary parts of Eq.~(9); and 3) we solve 
the ten equations and thus arrive at the analytical expressions of those ten parameters.
Different from the analytical results obtained previously in Refs.~\cite{Casas:1999tg,Antusch:2003kp}, our results are of the {\it integral}
form instead of the {\it differential} form. That is why we can express our 
results (mostly) in terms of the low-energy parameters by simply ascribing the 
RGE-induced running effects to the one-loop evolution parameter $\Delta^{}_\tau$,
making the issue much simpler and more transparent from the phenomenological point of view. 

First, the analytical results for the two initially nonzero neutrino masses are
\begin{eqnarray}
m^{}_2 (\Lambda^{}_{\rm F}) \simeq \hspace{-0.6cm}&& I^{}_0 \left[ 1 + \Delta^{}_\tau
\left( 2\cos^2{\theta^{}_{12}} \sin^2{\theta^{}_{23}} + 2\sin^2{\theta^{}_{12}}
\sin^2{\theta^{}_{13}} \cos^2{\theta^{}_{23}} \right.\right. \hspace{0.3cm}
\nonumber
\\
&& + \left.\left. \sin{2\theta^{}_{12}} \sin{\theta^{}_{13}} \sin{2\theta^{}_{23}}
\cos{\delta} \right) \right] m^{}_2 (\Lambda) \;,
\nonumber
\\
m^{}_3 (\Lambda^{}_{\rm F}) \simeq \hspace{-0.6cm}&& I^{}_0 \left( 1 + 2\Delta^{}_\tau
\cos^2{\theta^{}_{13}} \cos^2{\theta^{}_{23}}  \right) m^{}_3 (\Lambda)\;
\end{eqnarray}
in the $m^{}_1 (\Lambda) = 0$ case; or
\begin{eqnarray}
m^{}_1 (\Lambda^{}_{\rm F}) \simeq \hspace{-0.6cm}&& I^{}_0 \left[ 1 + \Delta^{}_\tau
\left( 2\sin^2{\theta^{}_{12}} \sin^2{\theta^{}_{23}} + 2\cos^2{\theta^{}_{12}}
\sin^2{\theta^{}_{13}} \cos^2{\theta^{}_{23}} \right.\right.
\nonumber
\\
&& - \left.\left. \sin{2\theta^{}_{12}} \sin{\theta^{}_{13}} \sin{2\theta^{}_{23}}
\cos{\delta} \right) \right] m^{}_1 (\Lambda) \;,
\nonumber
\\
m^{}_2 (\Lambda^{}_{\rm F}) \simeq \hspace{-0.6cm}&& I^{}_0 \left[ 1 + \Delta^{}_\tau
\left( 2\cos^2{\theta^{}_{12}} \sin^2{\theta^{}_{23}} + 2\sin^2{\theta^{}_{12}}
\sin^2{\theta^{}_{13}} \cos^2{\theta^{}_{23}} \right.\right. \hspace{0.3cm}
\nonumber
\\
&& + \left.\left. \sin{2\theta^{}_{12}} \sin{\theta^{}_{13}} \sin{2\theta^{}_{23}}
\cos{\delta} \right) \right] m^{}_2 (\Lambda) \;
\end{eqnarray}
in the $m^{}_3 (\Lambda) = 0$ case, where the flavor mixing angles $\theta^{}_{ij}$
(for $ij = 12, 13, 23$) and the CP-violating phase $\delta$ are all defined at
$\Lambda^{}_{\rm F}$.

As for the evolution of three lepton flavor mixing angles from $\Lambda$ down to
$\Lambda^{}_{\rm F}$, we have defined $\Delta \theta^{}_{ij} \equiv \theta^{}_{ij}
(\Lambda^{}_{\rm F}) - \theta^{}_{ij} (\Lambda)$ (for $ij = 12, 13, 23$) in Eq.~(14)
to describe the RGE-induced effects between the two energy scales. Our
one-loop analytical results are
\begin{eqnarray}
\Delta \theta^{}_{12} \simeq \hspace{-0.6cm}&& \frac{\Delta^{}_\tau}{2}
\left\{ \sin{2\theta^{}_{12}} \sin^2{\theta^{}_{13}} \cos^2{\theta^{}_{23}}
\left[ \zeta^{}_{32} \sin^2{ \left( \delta + \sigma \right) } + \zeta^{-1}_{32}
\cos^2{\left( \delta + \sigma \right)} - 1  \right] - \left[ \left(
\sin^2{\theta^{}_{23}} \right.\right.\right.
\nonumber
\\
&&- \left.\left. \sin^2{\theta^{}_{13}} \cos^2{\theta^{}_{23}} \right)
\sin{2\theta^{}_{12}} - \sin{\theta^{}_{13}}\sin{2\theta^{}_{23}}
\cos{2\theta^{}_{12}} \cos{\delta} \right] +  \sin{2\theta^{}_{23}}
\sin{\theta^{}_{13}}
\nonumber
\\
&& \times \left. \left[ \sin^2{\theta^{}_{12}} \cos{\delta}
+ \cos^2{\theta^{}_{12}} \left( \zeta^{}_{32} \sin{(\delta+\sigma)}\sin{\sigma}
+ \zeta^{-1}_{32} \cos{(\delta+\sigma)}\cos{\sigma} \right) \right] \right\} \;,
\nonumber
\\
\Delta \theta^{}_{13} \simeq \hspace{-0.6cm}&& -\frac{\Delta^{}_\tau}{2}
\left\{ \frac{1}{2} \sin{2\theta^{}_{12}} \sin{2\theta^{}_{23}}
\cos{\theta^{}_{13}} \left[ \zeta^{}_{32} \sin{\left( \delta + \sigma
\right)} \sin{\sigma} + \zeta^{-1}_{32} \cos{\left( \delta + \sigma
\right)} \cos{\sigma} - \cos{\delta} \right] \right. \hspace{0.4cm}
\nonumber
\\
&& + \left. \sin{2\theta^{}_{13}} \cos^2{\theta^{}_{23}} \left[ \left(
\zeta^{}_{32} \sin^2{\left( \delta + \sigma \right)} + \zeta^{-1}_{32}
\cos^2{\left( \delta + \sigma \right)} \right) \sin^2{\theta^{}_{12}}
+  \cos^2{\theta^{}_{12}} \right] \vphantom{\frac{1}{1}} \right\} \;,
\nonumber
\\
\Delta \theta^{}_{23} \simeq \hspace{-0.6cm}&& -\frac{\Delta^{}_\tau}{2}
\left\{ \sin{2\theta^{}_{12}} \sin{\theta^{}_{13}} \cos^2{\theta^{}_{23}}
\left[ \zeta^{}_{32} \sin{\left( \delta + \sigma \right)} \sin{\sigma} +
\zeta^{-1}_{32} \cos{\left( \delta + \sigma \right)} \cos{\sigma} -
\cos{\delta} \right] \right.
\nonumber
\\
&& + \left. \sin{2\theta^{}_{23}} \left[ \left( \zeta^{}_{32} \sin^2{\sigma}
+ \zeta^{-1}_{32} \cos^2{\sigma} \right) \cos^2{\theta^{}_{12}} +
\sin^2{\theta^{}_{12}} \right] \right\} \;
\end{eqnarray}
in the $m^{}_1 (\Lambda) = 0$ case; or
\begin{eqnarray}
\Delta \theta^{}_{12} \simeq \hspace{-0.6cm}&& - \frac{\Delta^{}_\tau}{2}
\left\{  \sin{2\theta^{}_{23}} \sin{\theta^{}_{13}} \left[ \cos{\delta} +
\frac{1}{2} \left( \zeta^{}_{21} -\zeta^{-1}_{21} \right) \sin{2\sigma}
\sin{\delta}  \right] + \left( \zeta^{}_{21} \sin^2{\sigma} + \zeta^{-1}_{21}
\cos^2{\sigma} \right) \right. \hspace{0.3cm}
\nonumber
\\
&& \times \left. \left[ \left( \sin^2{\theta^{}_{23}} - \sin^2{\theta^{}_{13}}
\cos^2{\theta^{}_{23}} \right) \sin{2\theta^{}_{12}} - \sin{\theta^{}_{13}}
\sin{2\theta^{}_{23}} \cos{2\theta^{}_{12}} \cos{\delta} \right]
\vphantom{\frac{1}{1}} \right\} \;,
\nonumber
\\
\Delta \theta^{}_{13} \simeq \hspace{-0.6cm}&& \frac{\Delta^{}_\tau}{2}
\sin{2\theta^{}_{13}} \cos^2{\theta^{}_{23}}   \;,
\nonumber
\\
\Delta \theta^{}_{23} \simeq \hspace{-0.6cm}&& \frac{\Delta^{}_\tau}{2}
\sin{2\theta^{}_{23}}  \;
\end{eqnarray}
in the $m^{}_3 (\Lambda) = 0$ case, where we have defined
$\zeta^{}_{ij} \equiv \left( m^{}_i - m^{}_j \right)/\left( m^{}_i + m^{}_j \right)$
with $m^{}_i$ and $m^{}_j$ being the neutrino masses at $\Lambda^{}_{\rm F}$
(for $i \neq j$ and $i,j =1,2,3$).

At the one-loop level it is well known that $m^{}_1 = 0$ (or $m^{}_3 = 0$) will
keep unchanged during the RGE running from $\Lambda$ to $\Lambda^{}_{\rm F}$,
and hence the corresponding Majorana CP phase $\rho$ (or $\varrho$) is not well
defined. In this case we only pay attention to the evolution of the remaining
two CP-violating phases $\delta$ and $\sigma$ by calculating $\Delta \delta \equiv
\delta (\Lambda^{}_{\rm F}) - \delta (\Lambda)$ and $\Delta \sigma \equiv
\sigma (\Lambda^{}_{\rm F}) - \sigma (\Lambda)$. Their approximate analytical
expressions turn out to be
\begin{eqnarray}
\Delta \delta \simeq \hspace{-0.6cm}&& \frac{\Delta^{}_\tau}{2} \left\{
\vphantom{\frac{1}{1}} \sin{2 \theta^{}_{12}} \sin{\theta^{}_{13}}
\cos{2\theta^{}_{23}} \cot{\theta^{}_{23}} \left[ \zeta^{}_{32}
\sin{\left( \delta + \sigma \right)} \cos{\sigma} - \zeta^{-1}_{32}
\cos{\left( \delta +\sigma \right)} \sin{\sigma} - \sin{\delta} \right]
\right.
\nonumber
\\
&& - \frac{2\sin{\theta^{}_{13}} \sin{2\theta^{}_{23}} \sin{\delta}}
{\sin{2\theta^{}_{12}}}  -\sin{2\theta^{}_{23}} \sin{\delta} \left(
\frac{\sin{2\theta^{}_{12}}}{2\sin{\theta^{}_{13}}} - \frac{2\sin{\theta^{}_{13}}}
{\sin{2\theta^{}_{12}}} \sin^4{\theta^{}_{12}} \right)- \sin{2\theta^{}_{23}}
\left( \frac{\sin{2\theta^{}_{12}}}{2\sin{\theta^{}_{13}}}  \right.
\nonumber
\\
&& - \left. \frac{2\sin{\theta^{}_{13}}}{\sin{2\theta^{}_{12}}}
\cos^4{\theta^{}_{12}} \right) \left[ \zeta^{}_{32} \cos{\left( \delta +
\sigma \right)} \sin{\sigma} - \zeta^{-1}_{32} \sin{\left( \delta + \sigma
\right)} \cos{\sigma} \right] + \left( \zeta^{}_{32} - \zeta^{-1}_{32} \right)
\nonumber
\\
&& \times \left. \left[ \left( \cos^2{\theta^{}_{12}} \sin^2{\theta^{}_{13}}
- \sin^2{\theta^{}_{12}} \right) \cos^2{\theta^{}_{23} \sin{2\left( \delta +
\sigma \right)}} + \cos^2{\theta^{}_{12}} \cos{2\theta^{}_{23}} \sin{2\sigma}
\right] \vphantom{\frac{1}{1}} \right\} \;
\end{eqnarray}
and
\begin{eqnarray}
\Delta \sigma  \simeq \hspace{-0.6cm}&& \frac{\Delta^{}_\tau}{2} \left\{
\sin{2\theta^{}_{12}} \sin{\theta^{}_{13}} \cot{\theta^{}_{23}} \left[
\sin{\delta} - \zeta^{}_{32} \sin{\left( \delta + \sigma \right)} \cos{\sigma}
+ \zeta^{-1}_{32} \cos{\left( \delta + \sigma \right)} \sin{\sigma} \right]
\right. \hspace{0.4cm}
\nonumber
\\
&&  + \left( \zeta^{}_{32} - \zeta^{-1}_{32} \right) \left[ \sin{2\theta^{}_{12}}
\sin{2\theta^{}_{23}} \sin{\theta^{}_{13}} \sin{\left( \delta + 2\sigma \right)}
- \cos^2{\theta^{}_{12}} \cos{2\theta^{}_{23}} \sin{2\sigma} \right.
\nonumber
\\
&&+ \left.\left. 2\sin^2{\theta^{}_{12}} \sin^2{\theta^{}_{13}}
\cos^2{\theta^{}_{23}} \sin{2\left( \delta + \sigma \right)} \right] \right\} \;
\end{eqnarray}
in the $m^{}_1 (\Lambda) = 0$ case; or
\begin{eqnarray}
\Delta \delta \simeq \hspace{-0.6cm}&& - \frac{\Delta^{}_\tau}{2} \left[
\vphantom{\frac{1}{1}} \left( \zeta^{}_{21} - \zeta^{-1}_{21} \right)
\sin{2\sigma} \left( \sin^2{\theta^{}_{23}} - \sin^2{\theta^{}_{13}}
\cos^2{\theta^{}_{23}} - \sin{\theta^{}_{13}} \sin{2\theta^{}_{23}}
\cot{2\theta^{}_{12}} \cos{\delta} \right) \right. \hspace{0.4cm}
\nonumber
\\
&& + \left. \frac{2\sin{\theta^{}_{13}} \sin{2\theta^{}_{23}} \sin{\delta}}
{\sin{2\theta^{}_{12}}} \left( \zeta^{}_{21} \cos^2{\sigma} + \zeta^{-1}_{21}
\sin^2{\sigma} \right) - 2\sin{\theta^{}_{13}} \sin{2\theta^{}_{23}}
\cot{2\theta^{}_{12}} \sin{\delta} \right] \;
\end{eqnarray}
and
\begin{eqnarray}
\Delta \sigma  \simeq \hspace{-0.6cm}&& - \frac{\Delta^{}_\tau}{2} \left\{
2 \sin{\theta^{}_{13}} \sin{2\theta^{}_{23}}
\sin{\delta} \left[ \left( \zeta^{}_{21} \cos^2{\sigma} + \zeta^{-1}_{21}
\sin^2{\sigma} \right) \cot{2\theta^{}_{12}} - \csc{2\theta^{}_{12}} \right]
\right. \hspace{0.4cm}
\nonumber
\\
&& + \left. \left( \zeta^{}_{21} -
\zeta^{-1}_{21} \right) \sin{2\sigma} \left[ \left( \sin^2{\theta^{}_{23}}
- \sin^2{\theta^{}_{13}} \cos^2{\theta^{}_{23}} \right) \cos{2\theta^{}_{12}}
\right.\right.
\nonumber
\\
&& - \left.\left. \sin{\theta^{}_{13}} \sin{2\theta^{}_{23}} \cos{2\theta^{}_{12}}
\cot{2\theta^{}_{12}} \cos{\delta}  \right] \right\} \;
\end{eqnarray}
in the $m^{}_3 (\Lambda) = 0$ case
\footnote{One should keep in mind that the unphysical phases $\phi^{}_\alpha$
(for $\alpha = e, \mu, \tau$) and $\rho$ (or $\varrho$) at the one-loop level
will also evolve with the energy scale $\mu$, and hence their evolution cannot be
ignored in deriving the one-loop RGEs of those physical flavor parameters
\cite{Mei:2003gn,Chankowski:1993tx,Babu:1993qv,
Antusch:2001ck,Antusch:2005gp,Mei:2005qp,Ohlsson:2013xva}.}.
These integral-form analytical results are new, and they are certainly  more
instructive and transparent than the differential RGEs of the relevant flavor
parameters for our understanding of their evolution behaviors from $\Lambda$
to $\Lambda^{}_{\rm F}$ at the one-loop level.
\begin{figure}[t]
  \centering
  \includegraphics[width=1\linewidth]{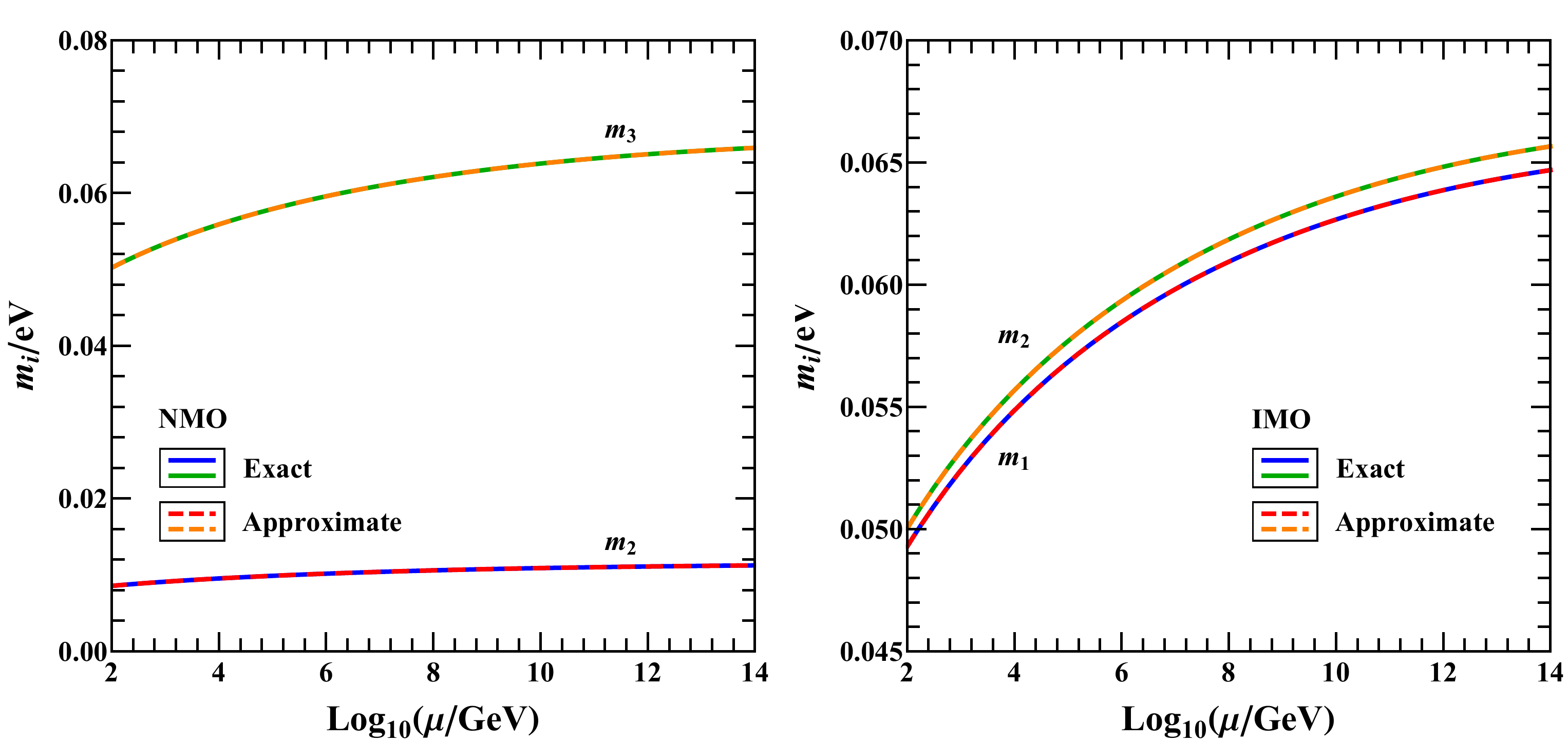}
  \caption{The numerical results of $m^{}_2$ and $m^{}_3$ (or $m^{}_1$ and
  $m^{}_2$) at a given energy scale above $\Lambda^{}_{\rm F}$ in the normal
  (or inverted) neutrino mass ordering case with $m^{}_1 = 0$ (or $m^{}_3 = 0$)
  at $\Lambda \simeq 10^{14}$ GeV and $\sigma (\Lambda^{}_{\rm F}) = 0$, $\pi/4$
  or $\pi/2$.}
  \label{m_i}
\end{figure}
\begin{figure}[t]
  \centering
  \includegraphics[width=0.99\linewidth]{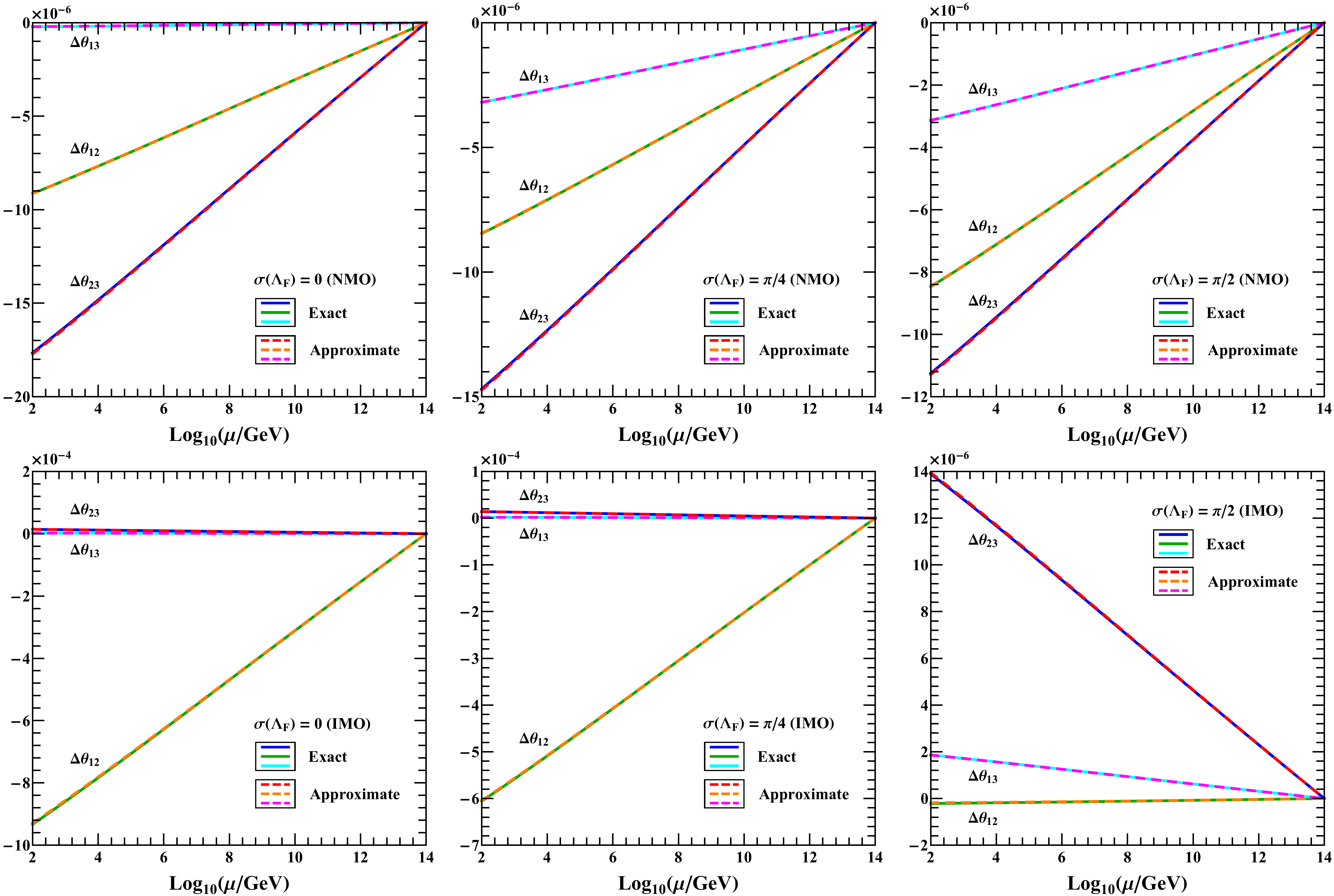}
  \caption{The numerical results of $\Delta\theta^{}_{ij}$ (for $ij=12,13,23$)
  at a given energy scale above $\Lambda^{}_{\rm F}$ in the normal (or inverted)
  neutrino mass ordering case with $m^{}_1 = 0$ (or $m^{}_3 = 0$) at
  $\Lambda \simeq 10^{14}$ GeV and $\sigma (\Lambda^{}_{\rm F}) = 0$, $\pi/4$
  or $\pi/2$.}
  \label{theta}
\end{figure}
\begin{figure}[t]
  \centering
  \includegraphics[width=0.99\linewidth]{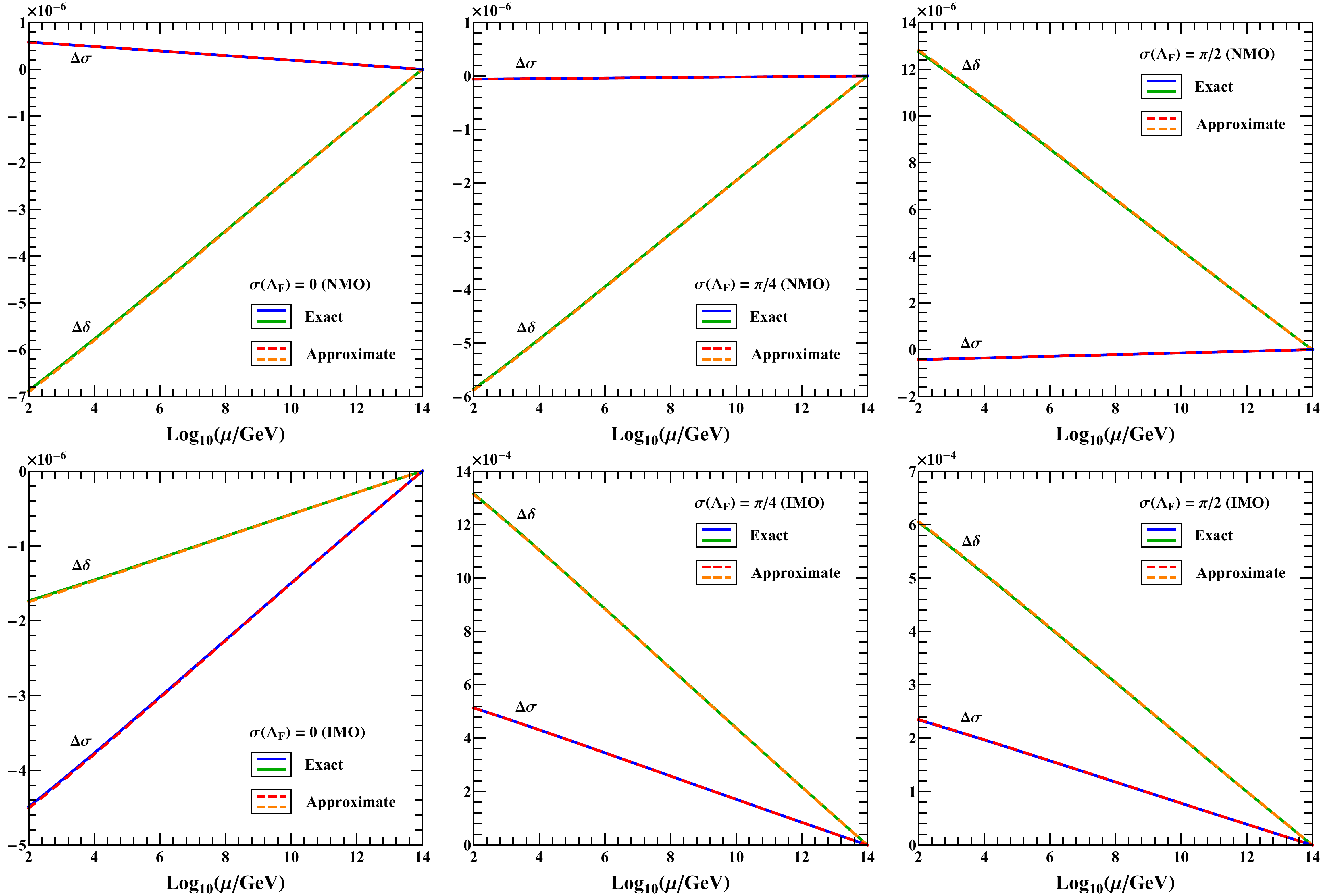}
  \caption{The numerical results of $\Delta\delta$ and $\Delta\sigma$
  at a given energy scale above $\Lambda^{}_{\rm F}$ in the normal (or inverted)
  neutrino mass ordering case with $m^{}_1 = 0$ (or $m^{}_3 = 0$) at
  $\Lambda \simeq 10^{14}$ GeV and $\sigma (\Lambda^{}_{\rm F}) = 0$, $\pi/4$
  or $\pi/2$.}
  \label{phase}
\end{figure}
\begin{table}[t!]
\centering
\caption{The values of $m^{}_2$ and $m^{}_3$ (or $m^{}_1$ and $m^{}_2$)
at $\Lambda^{}_{\rm F} \simeq 10^2$ GeV, together with those of $\Delta\theta^{}_{ij}$
(for $ij=12,13,23$), $\Delta\delta$ and $\Delta\sigma$, in the normal
(or inverted) neutrino mass ordering case with $m^{}_1 =0$ (or $m^{}_3 =0$)
at $\Lambda \simeq 10^{14}$ GeV and $\sigma (\Lambda^{}_{\rm F}) = 0$, $\pi/4$ or $\pi/2$.
The corresponding results given in the parentheses are obtained by numerically solving
the two-loop RGEs.}
\vspace{0.3cm}
\begin{tabular}{c|cccc}
\hline\hline
 & $\sigma \left( \Lambda^{}_{\rm F} \right)$ & $0$ & $\pi/4$ & $\pi/2$ \\
\hline
\multirow{12}{*}{NMO} & $m^{}_2 \left( \Lambda^{}_{\rm F} \right)/m^{}_2 \left( \Lambda \right)$
& 0.762 (0.762) & 0.762 (0.762) & 0.762 (0.762) \\
& $m^{}_3 \left( \Lambda^{}_{\rm F} \right)/m^{}_3 \left( \Lambda \right)$
& 0.762 (0.762) & 0.762 (0.762) & 0.762 (0.762) \\
& \multirow{2}{*}{$\Delta\theta^{}_{12} \left( \Lambda^{}_{\rm F} \right)$}
& $-9.124 \times 10^{-6}$ & $-8.445 \times 10^{-6}$ & $-8.459 \times 10^{-6}$  \\&
& ($-9.136 \times 10^{-6}$) & ($-8.460 \times 10^{-6}$) & ($-8.474 \times 10^{-6}$) \\
& \multirow{2}{*}{$\Delta\theta^{}_{13} \left( \Lambda^{}_{\rm F} \right)$}
& $-2.092 \times 10^{-7}$ & $-3.193 \times 10^{-6}$ & $-3.131 \times 10^{-6}$ \\&
& ($-2.227 \times 10^{-7}$) & ($-3.195 \times 10^{-6}$) & ($-3.135 \times 10^{-6}$) \\
& \multirow{2}{*}{$\Delta\theta^{}_{23} \left( \Lambda^{}_{\rm F} \right)$}
& $-1.772 \times 10^{-5}$ & $-1.474 \times 10^{-5}$ & $-1.129 \times 10^{-5}$ \\&
& ($-1.765 \times 10^{-5}$) & ($-1.469 \times 10^{-5}$) & ($-1.125 \times 10^{-5}$) \\
& \multirow{2}{*}{$\Delta\delta \left( \Lambda^{}_{\rm F} \right)$}
& $-6.902 \times 10^{-6}$ & $-5.882 \times 10^{-6}$ & $1.281 \times 10^{-5}$ \\&
& ($-6.869 \times 10^{-6}$) & ($-5.863 \times 10^{-6}$) & ($1.275 \times 10^{-5}$) \\
& \multirow{2}{*}{$\Delta\sigma \left( \Lambda^{}_{\rm F} \right)$}
& $5.829 \times 10^{-7}$ & $-5.920 \times 10^{-8}$ & $-4.205 \times 10^{-7}$ \\&
& ($5.807 \times 10^{-7}$) & ($-5.876 \times 10^{-8}$) & ($-4.190 \times 10^{-7}$) \\
\hline
\multirow{12}{*}{IMO} & $m^{}_1 \left( \Lambda^{}_{\rm F} \right)/m^{}_1 \left( \Lambda \right)$
& 0.762 (0.762) & 0.762 (0.762) & 0.762 (0.762) \\
& $m^{}_2 \left( \Lambda^{}_{\rm F} \right)/m^{}_2 \left( \Lambda \right)$
& 0.762 (0.762) & 0.762 (0.762) & 0.762 (0.762) \\
& \multirow{2}{*}{$\Delta\theta^{}_{12} \left( \Lambda^{}_{\rm F} \right)$}
& $-9.291 \times 10^{-4}$ & $-6.042 \times 10^{-4}$ & $-1.822 \times 10^{-7}$  \\&
& ($-9.326 \times 10^{-4}$) & ($-6.057 \times 10^{-4}$) & ($-2.138 \times 10^{-7}$) \\
& \multirow{2}{*}{$\Delta\theta^{}_{13} \left( \Lambda^{}_{\rm F} \right)$}
& $1.858 \times 10^{-6}$ & $1.858 \times 10^{-6}$ & $1.858 \times 10^{-6}$ \\&
& ($1.866 \times 10^{-6}$) & ($1.866 \times 10^{-6}$) & ($1.866 \times 10^{-6}$) \\
& \multirow{2}{*}{$\Delta\theta^{}_{23} \left( \Lambda^{}_{\rm F} \right)$}
& $1.394 \times 10^{-5}$ & $1.394 \times 10^{-5}$ & $1.394 \times 10^{-5}$ \\&
& ($1.389 \times 10^{-5}$) & ($1.389 \times 10^{-5}$) & ($1.389 \times 10^{-5}$) \\
& \multirow{2}{*}{$\Delta\delta \left( \Lambda^{}_{\rm F} \right)$}
& $-1.748 \times 10^{-6}$ & $1.313 \times 10^{-3}$ & $6.055 \times 10^{-4}$ \\&
& ($-1.732 \times 10^{-6}$) & ($1.315 \times 10^{-3}$) & ($6.039 \times 10^{-4}$) \\
& \multirow{2}{*}{$\Delta\sigma \left( \Lambda^{}_{\rm F} \right)$}
& $-4.508 \times 10^{-6}$ & $5.133 \times 10^{-4}$ & $2.347 \times 10^{-4}$ \\&
& ($-4.488 \times 10^{-6}$) & ($5.135 \times 10^{-4}$) & ($2.341 \times 10^{-4}$) \\
\hline\hline
\end{tabular}
\label{ovl}
\end{table}

With the same inputs as summarized in section 2, the evolution of those
initially nonzero flavor parameters, including $m^{}_2$ and $m^{}_3$ (or $m^{}_1$ and
$m^{}_2$) in the normal (or inverted) neutrino mass ordering case, $\Delta\theta^{}_{ij}$
(for $ij=12,13,23$), $\Delta\delta$ and $\Delta\sigma$, is numerically calculated
with the help of both the two-loop differential RGEs and the analytical
approximations given in Eqs.~(23)---(30). Our numerical results are illustrated in
Figs.~\ref{m_i}---\ref{phase}. In particular, the values of
such flavor parameters at $\Lambda^{}_{\rm F}$ are explicitly listed in Table~\ref{ovl},
where the numbers shown in the parentheses are obtained by numerically solving
the two-loop RGEs. Some immediate comments are in order.
\begin{itemize}
  \item From Eqs.~(23) and (24), one can see that the running effects of
  $m^{}_2$ and $m^{}_3$ (or $m^{}_1$ and $m^{}_2$) in the normal (or inverted)
  neutrino mass ordering case are mainly governed by an overall factor $I^{}_0$
  whose values changing with $\mu$ are shown in Fig.~\ref{loop function}, and they
  are independent of the value of the Majorana CP phase
  $\sigma \left( \Lambda^{}_{\rm F} \right)$ in the leading-order approximation,
  as also illustrated in Fig.~\ref{m_i} and Table~\ref{ovl}.

  \item In comparison with Eq.~(25), Eq.~(26) is much simpler and thus makes it much
  easier to understand the running behaviors of $\Delta\theta^{}_{ij}$ (for
  $ij= 12,13,23$) in the inverted neutrino mass ordering case. With the best-fit
  values of $\theta^{}_{ij}$, $\delta$, $\delta m^2$ and $\Delta m^2$ given in
  Eqs.~(21) and (22), it is obvious that in the inverted neutrino mass ordering case
  the evolution of $\Delta \theta^{}_{13}$ and $\Delta \theta^{}_{23}$ is
  dominated by that of $\Delta^{}_\tau$ and independent of the value of
  $\sigma \left( \Lambda^{}_{\rm F} \right)$ in the leading-order approximation,
  as also shown in Fig.~\ref{theta} and Table~\ref{ovl}.

  \item Fig.~\ref{theta} and Table~\ref{ovl} show that the magnitude of
  $\Delta\theta^{}_{13}$ is strongly suppressed in the normal neutrino mass ordering
  case with $\sigma \left( \Lambda^{}_{\rm F} \right) = 0$, mainly because
  a large cancellation appears in the analytical expression of $\Delta\theta^{}_{13}$
  when $\sigma \left( \Lambda^{}_{\rm F} \right) = 0$ is taken. The magnitude
  of $\Delta \theta^{}_{12}$ is also suppressed in the inverted mass ordering case
  with $\sigma \left( \Lambda^{}_{\rm F} \right) = \pi/2$, simply because of the
  suppression caused by   the smallness of $\zeta^{}_{21}$ and $\theta^{}_{13}$ when
  $\sigma \left( \Lambda^{}_{\rm F} \right) = \pi/2$ is taken. In either situation the
  relative accuracy of our analytical approximations at $\Lambda^{}_{\rm F}$ becomes
  worse, and it reduces from the $\mathcal{O}$(1\textperthousand) level to the
  $\mathcal{O}(1\%)$ level. Of course, the value of $\Delta\theta^{}_{12}$ is largely
  enhanced in the inverted neutrino mass ordering case with
  $\sigma \left( \Lambda^{}_{\rm F} \right) = 0$ or $\pi/4$ as a result of the
  largeness of $\zeta^{-1}_{21}$, which can easily be seen in Eq.~(26).

  \item As can be seen from Fig.~\ref{phase} and Table~\ref{ovl}, the value of
  $\Delta\sigma$ in the normal neutrino mass ordering case is much smaller than that
  in the inverted mass ordering case. In the latter case with
  $\sigma \left( \Lambda^{}_{\rm F} \right) = \pi/4$ or $\pi/2$, the values of
  $\Delta\delta$ and $\Delta\sigma$ are largely enhanced thanks to the largeness
  of $\zeta^{-1}_{21}$. Such a feature is easily understandable with the help of
  Eqs.~(29) and (30).
\end{itemize}

\section{Summary}

Given two different neutrino mass-squared differences that have been determined
in a number of neutrino oscillation experiments, whether the lightest neutrino
$\nu^{}_1$ (or $\nu^{}_3$) can be exactly massless turns out to be an interesting
question in neutrino phenomenology. From the perspective of model building,
it is always possible to obtain $m^{}_1 =0$ (or $m^{}_3 =0$) at the tree level
if the flavor structure of the model is properly specified (e.g., in the minimal
seesaw model with only two right-handed neutrino states). Then the question
becomes whether such a massless neutrino can stay massless against quantum
corrections when the energy scale evolves from a superhigh scale $\Lambda$,
where the seesaw mechanism or flavor symmetry works, down to the Fermi scale
$\Lambda^{}_{\rm F}$. In the SM framework Davidson {\it et al} have given a
preliminary answer to this question by taking into account the two-loop RGE-induced
effects \cite{Davidson:2006tg}. Here we have carried out a further study of
this issue by paying attention to the two-loop radiative corrections to not
only the smallest neutrino mass $m^{}_1$ (or $m^{}_3$) but also the associated
Majorana CP phase $\rho$ (or $\varrho$).

In the present work both $m^{}_1$ (or $m^{}_3$) and $\rho$ (or $\varrho$) at an
arbitrary energy scale between $\Lambda^{}_{\rm F}$ and $\Lambda$ have been
analytically formulated at the two-loop level, and their magnitudes have been evaluated
both based on our analytical approximations and by numerically solving the two-loop RGEs.
We find that the numerical results obtained in these two ways are in good
agreement with each other. In particular, we have confirmed that a nonzero value of
$m^{}_1$ (or $m^{}_3$) of ${\cal O}(10^{-13})$ eV at $\Lambda^{}_{\rm F}$
can be generated from $m^{}_1 =0$ (or $m^{}_3 =0$) at $\Lambda \simeq 10^{14}$ GeV
via the two-loop quantum corrections in the SM, and found that $\rho$ (or $\varrho$)
may accordingly acquire an appreciable physical value at the same level.
As a nontrivial by-product, the evolution of all those initially nonzero flavor
parameters of massive neutrinos has been calculated both analytically and
numerically, by simply keeping their leading (i.e., one-loop) RGE-induced effects.

This study can therefore allow one to draw the conclusion that taking $m^{}_1 =0$
(or $m^{}_3 =0$) and switching off the associated Majorana CP phase $\rho$ (or
$\varrho$) is absolutely safe at low energies for the minimal type-I seesaw model
and some other neutrino mass models of this kind which naturally predict
$m^{}_1 =0$ (or $m^{}_3 =0$) at the tree level at a superhigh energy scale.

\section*{Acknowledgements}

One of us (Z.Z.X.) is indebted to Xiangdong Ji for his interesting comments on the
possibility of a massless neutrino, and to Shun Zhou for many useful discussions on
the same issue. This research work is partly supported by the National Natural
Science Foundation of China under grant No. 11775231 and grant No. 11835013.

\vspace{0.5cm}


\end{document}